\newcommand\apjcls{1}
\newcommand\aastexcls{2}
\newcommand\othercls{3}

\newcommand\papercls{\aastexcls}
\documentclass[tighten, times,twocolumn]{aastex62}  

\if\papercls \apjcls
\usepackage{apjfonts}
\else\if\papercls \othercls
\usepackage{epsfig}
\usepackage{margin}
\usepackage{times}
\fi\fi
\usepackage{ifthen}
\usepackage{natbib}
\usepackage{bm}
\usepackage{amssymb, amsmath}
\usepackage{appendix}
\usepackage{etoolbox}
\usepackage[T1]{fontenc}
\usepackage{paralist}
\usepackage{newtxtext,newtxmath}
\if\papercls \apjcls 
\newcommand\aas{\ref@jnl{AAS Meeting Abstracts}}
\newcommand\dps{\ref@jnl{AAS/DPS Meeting Abstracts}}
\newcommand\maps{\ref@jnl{MAPS}}
\else\if\papercls \othercls
\usepackage{astjnlabbrev-jh}
\fi\fi

\if\papercls \aastexcls
\hypersetup{citecolor=blue, 
            linkcolor=blue, 
            menucolor=blue, 
            urlcolor=blue}  
\else
\usepackage[
bookmarks=true,           
bookmarksnumbered=true,   
colorlinks=true,          
citecolor=blue,           
linkcolor=blue,           
menucolor=blue,           
urlcolor=blue,            
linkbordercolor={0 0 1},  
pdfborder={0 0 1},
frenchlinks=true]{hyperref}
\fi

\if\papercls \othercls

\else

\fi

\providecommand{\adsurl}[1]{\href{#1}{ADS}}

\makeatletter
\patchcmd{\NAT@citex}
  {\@citea\NAT@hyper@{%
     \NAT@nmfmt{\NAT@nm}%
     \hyper@natlinkbreak{\NAT@aysep\NAT@spacechar}{\@citeb\@extra@b@citeb}%
     \NAT@date}}
  {\@citea\NAT@nmfmt{\NAT@nm}%
   \NAT@aysep\NAT@spacechar\NAT@hyper@{\NAT@date}}{}{}

\patchcmd{\NAT@citex}
  {\@citea\NAT@hyper@{%
     \NAT@nmfmt{\NAT@nm}%
     \hyper@natlinkbreak{\NAT@spacechar\NAT@@open\if*#1*\else#1\NAT@spacechar\fi}%
       {\@citeb\@extra@b@citeb}%
     \NAT@date}}
  {\@citea\NAT@nmfmt{\NAT@nm}%
   \NAT@spacechar\NAT@@open\if*#1*\else#1\NAT@spacechar\fi\NAT@hyper@{\NAT@date}}
\makeatother

\makeatletter
\DeclareRobustCommand{\lowcase}[1]{\@lowcase#1\@nil}
\def\@lowcase#1\@nil{\if\relax#1\relax\else\MakeLowercase{#1}\fi}
\makeatother

\DeclareSymbolFont{UPM}{U}{eur}{m}{n}
\DeclareMathSymbol{\umu}{0}{UPM}{"16}
\let\oldumu=\umu
\renewcommand\umu{\ifmmode\oldumu\else\math{\oldumu}\fi}

\if\papercls \othercls

\else

\fi

\let\oldsim=\sim
\renewcommand\sim{\ifmmode\oldsim\else\math{\oldsim}\fi}
\let\oldpm=\pm
\renewcommand\pm{\ifmmode\oldpm\else\math{\oldpm}\fi}
\newcommand\by{\ifmmode\times\else\math{\times}\fi}

\newbox{\wdbox}
\renewcommand\c{\setbox\wdbox=\hbox{,}\hspace{\wd\wdbox}}
\renewcommand\i{\setbox\wdbox=\hbox{i}\hspace{\wd\wdbox}}

\newcount\timect
\newcount\hourct
\newcount\minct
\newcommand\now{\timect=\time \divide\timect by 60
         \hourct=\timect Cltiply\hourct by 60
         \minct=\time \advance\minct by -\hourct
         \number\timect:\ifnum \minct < 10 0\fi\number\minct}

\catcode`@=11

\newcommand\comment[1]{}

\newcommand\commenton{\catcode`\%=14}

\renewcommand\math[1]{$#1$}
\newcommand\mathshifton{\catcode`\$=3}

\let\atab=&
\newcommand\atabon{\catcode`\&=4}

\let\oldmsp=\sp
\let\oldmsb=\sb
\def\sp#1{\ifmmode
           \oldmsp{#1}%
         \else\strut\raise.85ex\hbox{\scriptsize #1}\fi}
\def\sb#1{\ifmmode
           \oldmsb{#1}%
         \else\strut\raise-.54ex\hbox{\scriptsize #1}\fi}
\newbox\@sp
\newbox\@sb
\def\sbp#1#2{\ifmmode%
           \oldmsb{#1}\oldmsp{#2}%
         \else
           \setbox\@sb=\hbox{\sb{#1}}%
           \setbox\@sp=\hbox{\sp{#2}}%
           \rlap{\copy\@sb}\copy\@sp
           \ifdim \wd\@sb >\wd\@sp
             \hskip -\wd\@sp \hskip \wd\@sb
           \fi
        \fi}
\def\msp#1{\ifmmode
           \oldmsp{#1}
         \else \math{\oldmsp{#1}}\fi}
\def\msb#1{\ifmmode
           \oldmsb{#1}
         \else \math{\oldmsb{#1}}\fi}

\def\supon{\catcode`\^=7}

\def\subon{\catcode`\_=8}

\def\supsubon{\supon \subon}

\newcommand\actcharon{\catcode`\~=13}

\newcommand\paramon{\catcode`\#=6}

\comment{And now to turn us totally on and off...}

\newcommand\reservedcharson{ \commenton  \mathshifton  \atabon  \supsubon 
                             \actcharon  \paramon}

\catcode`@=12
\reservedcharson

\if\papercls \apjcls

\else

\fi

\newcommand\chisq{\ifmmode{\chi\sp{2}}\else\math{\chi\sp{2}}\fi}
\newcommand\redchisq{\ifmmode{ \chi\sp{2}\sb{\rm red}}
                    \else\math{\chi\sp{2}\sb{\rm red}}\fi}
\newcommand\Teq{\ifmmode{T\sb{\rm eq}}\else$T$\sb{eq}\fi}
\newcommand\mjup{\ifmmode{M\sb{\rm Jup}}\else$M$\sb{Jup}\fi}
\newcommand\rjup{\ifmmode{R\sb{\rm Jup}}\else$R$\sb{Jup}\fi}
\newcommand\msun{\ifmmode{M\sb{\odot}}\else$M\sb{\odot}$\fi}
\newcommand\rsun{\ifmmode{R\sb{\odot}}\else$R\sb{\odot}$\fi}
\newcommand\mearth{\ifmmode{M\sb{\oplus}}\else$M\sb{\oplus}$\fi}
\newcommand\rearth{\ifmmode{R\sb{\oplus}}\else$R\sb{\oplus}$\fi}

\shorttitle{Fast and Slow Crystallization-driven Convection in White Dwarfs}
\shortauthors{M. Castro-Tapia {\em et al.}}

\begin{document}

\title{Fast and Slow Crystallization-driven Convection in White Dwarfs}
\author{Matias Castro-Tapia}
\affiliation{\rm Department of Physics and Trottier Space Institute, McGill University, Montreal, QC H3A 2T8, Canada}

\author{Andrew Cumming}
\affiliation{\rm Department of Physics and Trottier Space Institute, McGill University, Montreal, QC H3A 2T8, Canada}

\author{J. R. Fuentes}
\affiliation{\rm Department of Applied Mathematics, University of Colorado Boulder, Boulder, CO 80309-0526, USA}


\begin{abstract}

We investigate crystallization-driven convection in carbon–oxygen white dwarfs. We present a version of the mixing length theory that self-consistently includes the effects of thermal diffusion and composition gradients, and provides solutions for the convective parameters based on the local heat and composition fluxes. Our formulation smoothly transitions between the regimes of fast adiabatic convection at large Peclet number and slow thermohaline convection at low Peclet number. It also allows for both thermally driven and compositionally driven convection, including correctly accounting for the direction of heat transport for compositionally driven convection in a thermally stable background. We use the MESA stellar evolution code to calculate the composition and heat fluxes during crystallization in different models of cooling white dwarfs, and determine the regime of convection and the convective velocity. We find that convection occurs in the regime of slow thermohaline convection during most of the cooling history of the star. However, at the onset of crystallization, the composition flux is large enough to drive fast overturning convection for a short time ($\sim$10 Myr). We estimate the convective velocities in both of these phases and discuss the implications for explaining observed white dwarf magnetic fields with crystallization-driven dynamos.
\end{abstract}

\keywords{stars: interiors, stars: convection, white dwarfs}

\section{Introduction}
As white dwarfs (WDs) cool over time, a first-order phase transition leads to the solidification of their degenerate interiors \citep{vanHorn1968}. This crystallization process occurs from the core outward. In carbon–oxygen WDs, the transition from liquid to solid preferentially retains oxygen in the solid core, creating an unstable, buoyant region on top of it as lower-density carbon is released. Thus, convection is induced as the crystallization front grows outward, changing the composition profile \citep{Stevenson1980, Mochkovitch1983, Isern1997, Fuentes2023}. The latent heat released during the phase transition can significantly slow down the cooling of the star, and the delay is increased even more by the gravitational energy released as the elements are redistributed. The cooling delay leads to an overdensity in the Hertzsprung–Russell diagram, known as the Q-branch, that has been revealed by Gaia \citep{Tremblay2019}, although an additional source of cooling delay seems to be required to explain the observed overdensity \citep{Cheng2019}.

The crystallization process has also been linked to magnetism in white dwarfs. It has been suggested that compositionally driven convection during crystallization can drive a dynamo \citep[e.g.,][]{Isern2017, Ginzburg2022}. This would explain why most magnetic white dwarfs appear at low temperatures and luminosities \citep[e.g.,][]{Liebert2003, Sion2014} as well as their incidence in different close binary systems \citep{Belloni2021, Schreiber2021, Schreiber2022}. However, whether convection is efficient enough to explain the large intensity of the observed magnetic fields ($B_{\mathrm{obs}}\sim0.1$--$1000$ MG) is still under debate. If convection is efficient and the gravitational energy released during crystallization is used to power the magnetic field, \citet{Isern2017} showed that the saturated dynamo scalings of \citet{Christensen2009} \citep[see also][]{Christensen2010} can explain field strengths in the MG range. Taking rotation into account could potentially explain larger fields \citep{Isern2017, Ginzburg2022}.

Although the energetics of a crystallization-driven dynamo seem reasonable, it is not clear whether the convective velocity is large enough.For a saturated dynamo, we expect $\rho v_c^2\sim B^2/4\pi$, which gives $v_c\sim 100\ {\rm cm\ s^{-1}}$ for $B\sim 1\ \mathrm{MG}$ \citep{Ginzburg2022}. However, this velocity is much larger than expected based on estimates from mixing length theory (MLT) applied to white dwarfs, in which the large thermal conductivity reduces the convective efficiency by reducing the effective buoyancy of rising fluid elements \citep{Mochkovitch1983, Isern1997}. The kinetic energy flux is then a small fraction of the total energy flux. \citet{Mochkovitch1983} estimated convective velocities in the range $10^{-6}$--$10^{-1}~\mathrm{cm\ s^{-1}}$, depending on the star's rotation rate. For such small velocities, if the kinetic energy carried by convection powers the magnetic field, the intensity of the field is restricted to a few Gauss, much lower than observed.

Using both MLT and hydrodynamic simulations, we recently showed that there are two modes of compositionally driven convection, depending on the magnitude of the composition flux (\citealt{Fuentes2023}, hereafter F23). When the composition flux is larger than some threshold, convection occurs in a fast mode in which the convective motions are too rapid for thermal diffusion to be important. For this fast mode of convection, the velocity estimates are similar to those of \citet{Ginzburg2022}. Below the threshold composition flux, convection is slow, with thermal diffusion acting to reduce the convective velocities to the small values found by \citet{Mochkovitch1983}. As a first estimate, F23 used cooling models of white dwarfs to obtain the composition flux, finding that, except for a brief period at the onset of crystallization, convection in white dwarfs occurs in the slow mode, with correspondingly small convective velocities. However, these estimates were approximate since the models used did not include phase separation self-consistently, and F23 assumed that the inward convective and outward conductive heat fluxes exactly balance (zero net heat flux).

Current evolution models of cooling WDs that include crystallization do not explicitly follow convection. Instead, it is usually assumed that the liquid phase above the core is fully mixed, because the relevant timescales considered for the evolution are much larger than the convective timescale (see, e.g., ~\citealt{Althaus2012} for LPCODE, \citealt{Salaris2022} for BaSTI, \citealt{Bauer2023} for MESA). The energy contribution from the redistribution of chemical elements is then added to the structure equations as a source term following the prescriptions in \citet{Isern1997, Isern2000}. The energy released by the phase transition itself, i.e., the latent heat, is naturally included as a source term in the local solidification front. Such a treatment simplifies the computation of the cooling while still including the effect of phase transition on the total energy released and therefore cooling delay. However, the assumption that the layers above the solid core are instantaneously mixed means that the parameters of convection are not calculated.

One might imagine that, to follow convection in a cooling white dwarf, we could drop the assumption of instantaneous mixing and instead allow the convective mixing routine in the code to do the mixing for us. However, the standard prescriptions for convection in stellar evolution codes are not applicable to compositionally driven convection in white dwarfs. A prescription for inefficient convection is usually included in stellar evolution codes in which MLT is modified to include radiative losses (e.g., see \citealt{Kippenhahn2013} which is based on \citealt{BohmVitense1958}). However, this applies only for overturning convection in which the temperature gradient is superadiabatic (for example, in surface convection zones of stars). In the slow regime of convection in white dwarfs, thermal diffusion is efficient enough that the temperature gradient remains subadiabatic. This is the regime of thermohaline convection (e.g.,~\citealt{Kippenhahn1980, Kippenhahn2013}), which is typically modeled with a separate prescription that transports composition but not heat (e.g.,~\citealt{Kato1966, Kippenhahn1980, Langer1983}). Even in the fast regime of compositionally driven convection where thermal losses can be neglected, standard MLT implementations do not apply, since they usually assume that the convective heat transport is outward, whereas compositionally driven convection in white dwarfs transports heat inward (F23).

In this paper, we discuss the formalism needed to follow convection and the associated composition and heat transport in 1D cooling white dwarf models. We extend the MLT developed by F23, which assumed zero net heat flux, to arbitrary heat and composition fluxes. We show that the cubic equation that is usually solved to obtain the convective velocity in MLT is replaced by a fifth-order equation when composition gradients and thermal losses are correctly accounted for. A similar equation was derived by \citet{Umezu1988} for thermally driven overturning convection in massive stars; here, we extend this result to also include compositionally driven overturning convection and thermohaline convection.

We start in Section~\ref{sec:MLT} with a discussion of the standard implementations of MLT for inefficient convection and its implementation in stellar evolution codes and then present our formulation for studying compositionally driven convection.  In Section~\ref{sec:3} we explore the regimes of compositionally driven convection as a function of the total heat flux and composition flux. In Section~\ref{sec:WD}, we discuss the application of our theory to cooling models of WDs. In particular, we find that convection occurs in the fast regime for a short time after the onset of crystallization ($\lesssim 10\ \mathrm{Myr}$), but then quickly transitions and remains in the slow regime for the rest of the evolution. In Section~\ref{sec:summary_conclusions}, we conclude and discuss the implications for future white dwarf models and crystallization-driven dynamos.

\section{Mixing Length Theory}\label{sec:MLT}

In this section, we first review the standard treatment of inefficient convection in the framework of MLT (Section \ref{sec:2.1}), and then discuss the ways in which it has been extended to include composition gradients in different stellar evolution codes (Section \ref{sec:2.2}). Finally, we develop a general MLT that covers the regimes of thermohaline and overturning convection that we can apply to white dwarf interiors (Section \ref{sec:new_mlt}). Additionally, we show how our MLT prescription relates to previous implementations of thermohaline convection (Section \ref{sec_ineff_eff_conv})

\subsection{Inefficient Convection without Composition Gradients}\label{sec:2.1}

We first discuss the usual implementation of inefficient convection without composition gradients, following the prescription of \citet{CoxGiuli1968}.

The density contrast of convecting fluid elements is written as

\begin{equation}\label{eq_drho}
   \frac{D\rho}{\rho}=-\frac{\chi_{T}}{\chi_{\rho}}\frac{DT}{T}=-\frac{\chi_{T}}{\chi_{\rho}}(\nabla-\nabla_{e})\frac{\ell}{2H_{P}}~,
\end{equation}
where we have used the notation of \citet{Kippenhahn2013}: $D\rho$ and $DT$ are the density and temperature excess carried by a fluid element, respectively, $\ell$ is the mixing length, $H_{P}$ is the pressure scale height, $\chi_T = \left.\partial\ln P/\partial\ln T\right|_{\rho, X}$, and $\chi_\rho = \left.\partial\ln P/\partial\ln \rho\right|_{T, X}$. The subscript $X$ is used to represent the partial derivatives taken at constant composition. The temperature gradient with pressure in the star is $\nabla=d\ln{T}/d\ln{P}|_{\star}$ and the temperature gradient with pressure experienced by a fluid element as it moves is $\nabla_{e}$. The buoyancy force associated with the density excess $D\rho$ gives the convective velocity as\footnote{Note that we follow the standard treatment of MLT here that does not include rotation or magnetic forces; we discuss the possible effect of including these terms in Section \ref{sec:summary_conclusions}.} $v_c^{2}\approx -g(D\rho/\rho)(\ell/4)$ or

\begin{equation}\label{eq_vc}
    v_c=\sqrt{\frac{\chi_{T}}{\chi_{\rho}}\frac{g}{8H_{p}}}\ell(\nabla-\nabla_{e})^{1/2}~,
\end{equation}
where $g$ is the local gravitational acceleration. 

The effect of thermal diffusion is to set $\nabla_e$. If heat loss is negligible, fluid elements move adiabatically and $\nabla_e=\nabla_\mathrm{ad}$; if heat loss is very efficient, the fluid element experiences the same gradient as the background, $\nabla_e\rightarrow \nabla$. The value of $\nabla_e$ between these limits is determined by how quickly the fluid element moves. Considering the exchange of energy between the fluid element and the surroundings gives (\citealt{Kippenhahn2013} based on \citealt{BohmVitense1958})

\begin{equation}\label{eq_gamma}
    \Gamma =  \frac{\nabla-\nabla_{e}}{\nabla_{e}-\nabla_{\mathrm{ad}}} = \frac{1}{2a_{0}}\frac{v_{c}\ell}{\kappa_{T}},
\end{equation}
where $a_{0}$ is a geometric term that depends on the assumptions for the shape of the fluid element, and $\kappa_{T}=4acT^{3}/(3\kappa \rho^{2}c_{P})$ is the thermal diffusivity, where $\kappa$ is the opacity, $a$ the radiation constant, and $c$ the speed of light. The dimensionless quantity $\Gamma$ is usually taken as a measurement of the convective efficiency since it describes how closely convecting fluid elements follow the background gradient and therefore how easily they are accelerated by the buoyancy forces \citep{Jermyn2022}. 
Note that $\Gamma$ is proportional to the P\'eclet number $\mathrm{Pe}\equiv v_c\ell/\kappa_T$ (see Eq.~(4) of F23) which measures the ratio of thermal diffusion timescale to convective turnover time. The adiabatic limit (efficient convection) corresponds to $\Gamma\gg 1$, while inefficient convection occurs when $\Gamma\ll 1$.
Using Equation~\eqref{eq_vc} for the velocity gives

\begin{equation}\label{eq_gamma_2}
    \Gamma=A(\nabla-\nabla_{e})^{1/2},
\end{equation}
where we define

\begin{equation}
A=\frac{1}{2a_{0}}\sqrt{\frac{\chi_{T}}{\chi_{\rho}}\frac{g}{8H_{p}}}\frac{\ell^{2}}{\kappa_{T}}.
\end{equation}
Note that, for a constant value of $A$, we have $v_c\propto \Gamma\propto (\nabla-\nabla_{e})^{1/2}$. 

The convective heat flux is written as $F_{c}=\rho v_c Dq=\rho v_c T Ds$, where we relate the heat and entropy excess using the second law of thermodynamics $Dq = T\,Ds$. Also, as long as the fluid is compositionally homogeneous, $Ds = c_p DT/T$, where $c_{P}$ is the specific heat capacity at constant pressure, giving

\begin{equation}\label{eq_Fc}
    F_c = \rho v_c c_{P} DT = \rho v_c c_{P} T\left(\nabla-\nabla_e\right)\frac{\ell}{2H_{P}}~.
\end{equation}

The radiative heat flux is given in terms of the temperature gradient as

\begin{equation}\label{eq_Frad}
    F_{\mathrm{rad}}=\rho c_{P} \kappa_{T}\frac{T}{H_{P}}\nabla~.
\end{equation}
It is also useful to define the gradient $\nabla_\mathrm{rad}$ that measures the total heat flux, 

\begin{equation}\label{eq_Ftot}
 F=F_{\mathrm{rad}}+F_c=\rho c_{P} \kappa_{T}\frac{T}{H_{P}}\nabla_{\mathrm{rad}}~.
\end{equation}
Using Equation~\eqref{eq_vc} for the velocity, we obtain a relation between the temperature gradients 

\begin{equation}\label{eq_del_rad_del}
    \nabla_{\mathrm{rad}}-\nabla=a_{0}A(\nabla-\nabla_{e})^{3/2}~.
\end{equation}
Thus, when there are no composition gradients, the convective heat flux is proportional to $(\nabla-\nabla_{e})^{3/2}$.

If the total heat flux (given by $\nabla_{\mathrm{rad}}$) and the adiabatic gradient $\nabla_{\mathrm{ad}}$ 
are known quantities, then $\nabla$, $\nabla_{e}$, and $\Gamma$ can be found using Equations~\eqref{eq_gamma}, \eqref{eq_gamma_2} and \eqref{eq_del_rad_del} (given the constants $a_0$ and $A$). The solution for $\Gamma$ is the only real and positive root of the following third-degree polynomial equation\footnote{Equation \eqref{eq_poly} is sometimes written in terms of $\nabla$ rather than $\Gamma$, e.g.,~\cite{Maeder2009} and \cite{Kippenhahn2013}.} 

\begin{equation}\label{eq_poly}
    a_{0}\Gamma^{3}+\Gamma^{2}+\Gamma-A^{2}(\nabla_{\mathrm{rad}}-\nabla_{\mathrm{ad}})=0.
\end{equation}

Denoting the root as $\Gamma_{0}$, the solutions for the temperature gradients are

\begin{equation}\label{eq_del}
    \nabla=(1-\zeta)\nabla_{\mathrm{rad}}+\zeta\nabla_{\mathrm{ad}}~,
\end{equation}
\begin{equation}\label{eq_del_del_e}
    \nabla-\nabla_{e}=(\nabla_{\mathrm{rad}}-\nabla_{\mathrm{ad}})(1-\zeta)\frac{\Gamma_{0}}{1+\Gamma_{0}}~,
\end{equation}
where 

\begin{equation}\label{eq:zeta}
\zeta=\frac{a_{0}\Gamma_{0}^{2}}{1+\Gamma_{0}(1+a_{0}\Gamma_{0})}
\end{equation}
is another measure of the convective efficiency. 

MLT is typically implemented in stellar evolution codes by solving Equation \eqref{eq_poly} or its equivalent to determine the temperature gradient $\nabla$ given the total heat flux $\nabla_\mathrm{rad}$ (e.g., the \texttt{mlt} routine in MESA, \citealt{Paxton2011}). When $\zeta\rightarrow 1$, convection is efficient with $\Gamma\gg 1$ and the temperature gradient is close to adiabatic $\nabla\rightarrow \nabla_\mathrm{ad}$. When $\zeta\rightarrow 0$, convection is inefficient, $\Gamma\ll 1$, and radiation carries most of the heat with a superadiabatic gradient $\nabla\rightarrow \nabla_\mathrm{rad}$.

\subsection{Implementation of Composition Gradients in Stellar Evolution Codes}\label{sec:2.2}

Composition gradients modify the criterion for the onset of convection from the Schwarzschild criterion $\nabla>\nabla_{\mathrm{ad}}$ to the Ledoux criterion

\begin{equation}\label{eq_Ledoux}
    \nabla > \nabla_{\mathrm{ad}}-\nabla_{\mathrm{com}}= \nabla_L,
\end{equation}
 where $\nabla_{\mathrm{com}}=\chi_{T}^{-1}\sum_{i=1}^{n-1}\chi_{X_{i}}\nabla_{X_{i}}$ represents the composition gradients\footnote{In this notation, $\nabla_{\mathrm{com}}$ is equivalent to $-B$ in Equation (6) of \citet{Paxton2013}, where the implementation of composition gradients in MESA is described. This quantity is also usually written in terms of the mean molecular weight $\mu$, which simplifies the summation term, $\nabla_{\mathrm{com}}=-(\varphi/\delta)\nabla_{\mu}$ \citep[see][]{Maeder2009, Kippenhahn2013}, where $\varphi=\partial\ln\rho/\partial\ln\mu|_{P,T}$ and $\delta=-\partial\ln\rho/\partial\ln T|_{P,\mu}$.}, $\chi_{X_{i}} = \left.\partial\ln P/\partial\ln X_{i}\right|_{\rho, T, X_{j\neq i}}$, and  $\nabla_{X_{i}}=d \ln X_{i}/d \ln P|_{\star}$ for $n$ species with mass fractions $X_{i}$. The mass abundances must fulfill $\sum_{i=1}^{n}X_{i}=1$, which allows us to eliminate one of the abundances in the $\nabla_{\mathrm{com}}$ definition, whose summation has been written from 1 to $n-1$.
Depending on the sign of $\nabla_\mathrm{com}$, composition gradients can be either stabilizing ($\nabla_\mathrm{com}<0$) or destabilizing ($\nabla_\mathrm{com}>0$). 

When the temperature gradient needed to carry the heat flux by radiation $\nabla_\mathrm{rad}$ is unstable according to the Ledoux criterion, convection will occur and MLT can be applied. Different stellar evolution codes have taken different approaches to incorporating the term $\nabla_{\mathrm{com}}$ in the basic Equations ~\eqref{eq_gamma}, \eqref{eq_gamma_2}, and \eqref{eq_del_rad_del}.
For instance, in MESA, the definition of $\Gamma$ in Equation \eqref{eq_gamma} is modified by replacing the adiabatic gradient with $\nabla_{\mathrm{L}}$, while the expressions for convective velocity and flux (Eqs.~\eqref{eq_gamma_2} and ~\eqref{eq_del_rad_del}) are left unchanged (see \citealt{Jermyn2023} Equations~(18), (19) and (25) for flux, $\Gamma$, and $v_c$ respectively). This gives a convective flux $\propto (\Gamma/(\Gamma+1))^{3/2}(\nabla-\nabla_L)^{3/2}$.

In \cite{Vazan2015}, $\Gamma$ is defined as $(\nabla-\nabla_e+\nabla_\mathrm{com})/(\nabla_e-\nabla_\mathrm{ad})$, i.e.,~the numerator of $\Gamma$ is modified rather than the denominator. In addition, the composition gradient is included in the buoyancy force by writing $v_c\propto (\nabla-\nabla_e+\nabla_{\mathrm{com}})^{1/2}$ (see their eq.~(A10)) and in the convective heat flux by writing the heat flux in terms of the enthalpy excess $Dh$ as $F_{c}=\rho v_{c} Dh$. This leads to a scaling $F_c\propto v_c(\nabla-\nabla_e+\nabla_{\mathrm{com}})\propto (\nabla-\nabla_e+\nabla_{\mathrm{com}})^{3/2}$ (see Equation~(A12) of \citealt{Vazan2015}; also \citealt{Kuhfuss1986, Straka2005}).

As well as being different from each other, these implementations have a number of problems when applied to compositionally driven convection in white dwarfs. First, as we discuss in Appendix \ref{sec:appendix}, the composition terms in the entropy excess, which determines the heat flux, are a small correction, so that writing the heat flux in terms of the enthalpy excess overestimates the effect of the composition terms. Second, these implementations do not correctly handle the case where convection is driven by large destabilizing composition gradients that overcome a stable thermal gradient. When $F_c\propto (\nabla-\nabla_e+\nabla_{\mathrm{com}})^{3/2}$, the convective heat flux is always outward when the Ledoux criterion is satisfied, whereas convection should transport heat inward in a thermally stable background (F23). 
This implies that the equations being used are not internally self-consistent; indeed, we will see below that the cubic equation that is usually solved to obtain $\Gamma$ (e.g.~eq.~(A15) of \citealt{Vazan2015}) should be replaced by a quartic or quintic equation when composition gradients are included.

When the Ledoux criterion is not satisfied ($\nabla<\nabla_L$), the fluid is overall stable to convection and MLT gives $\Gamma=0$ (no convection). However, double-diffusive instabilities can still drive mixing if either the thermal gradient or the composition gradient is unstable. If the composition gradient is overall stabilizing, but the thermal gradient is unstable, this leads to semiconvection. In the case relevant for white dwarf crystallization, we have a stabilizing thermal gradient, but an unstable composition gradient, leading to thermohaline convection. These different kinds of double-diffusive convection are usually handled with a mixing prescription that is separate from MLT; see, e.g.,~\cite{Langer1983} for semiconvection or \cite{Kippenhahn1980}, \cite{Brown2013}, and \cite{Lattanzio2015} for thermohaline convection. In particular, for thermohaline convection, the mixing is usually done by defining an effective diffusion coefficient for composition and it is assumed that all the heat transport is by radiative diffusion. Therefore, in regions undergoing thermohaline convection, there is no consideration of the convective heat transport. In the next section, we present an extension of MLT that smoothly transitions between the thermohaline and overturning convection regimes.

\subsection{MLT Including Compositionally Driven Convection}\label{sec:new_mlt}

Having motivated the need for a more complete version of mixing length theory to model compositionally driven convection, we now extend the MLT derived by F23 to arbitrary heat and composition fluxes. Whereas F23 assumed that the inward convective heat flux and outward conductive flux were always in balance, our goal here is to generalize this so that, given the heat flux and composition flux, we can predict the temperature and composition gradients required for the stellar model.

We proceed as in Section~\ref{sec:2.1}, but including composition gradients in each step. The density contrast in the convection zone is

\begin{align}\label{eq_del_rho_comp}
    \nonumber\frac{D\rho}{\rho}&=-\frac{\chi_{T}}{\chi_{\rho}}\frac{DT}{T}-\frac{1}{\chi_{\rho}}\sum_{i=1}^{n-1}\chi_{X_{i}}\frac{DX_{i}}{X_{i}}~,\\
    &=-\frac{\chi_{T}}{\chi_{\rho}}(\nabla-\nabla_{e}+\nabla_{\mathrm{com}})\frac{\ell}{2H_{P}}~.
\end{align}
Since the buoyant acceleration is proportional to $-D\rho/\rho$, continued convection requires $D\rho<0$ or

\begin{equation}\label{eq_continuous_conv}
\nabla-\nabla_{e}+\nabla_{\mathrm{com}}>0.
\end{equation}
In the limit of rapid convection where thermal diffusion can be neglected, $\nabla_{e}\rightarrow\nabla_{\mathrm{ad}}$, and condition \eqref{eq_continuous_conv} is just the usual form of the Ledoux criterion (Equation~\eqref{eq_Ledoux}). In the case of slow thermohaline convection, however, thermal diffusion reduces $\nabla_e$ to a value well below $\nabla_\mathrm{ad}$, so that an unstable composition gradient $\nabla_\mathrm{com}>0$ can sustain convection even though the adiabatic Ledoux criterion is not satisfied.

The convective velocity is

\begin{equation}\label{eq_vc2}
    v_{c}=\sqrt{\frac{\chi_{T}}{\chi_{\rho}}\frac{g}{8H_{p}}}\ell(\nabla-\nabla_{e}+\nabla_{\mathrm{com}})^{1/2}~,
\end{equation}
differing from Equation~\eqref{eq_vc} due to the addition of $\nabla_{\mathrm{com}}$.
Using the definition of $\Gamma$ from Equation \eqref{eq_gamma}, we obtain the general form of the convective efficiency

\begin{equation}\label{eq_gamma_3}
    \Gamma=A(\nabla-\nabla_{e}+\nabla_{\mathrm{com}})^{1/2}~,
\end{equation}
which replaces Equation \eqref{eq_gamma_2}.
Equation~\eqref{eq_gamma} still applies even with composition gradients, since it can be obtained using just the temperature excesses \citep{Maeder2009, Kippenhahn2013}.
As mentioned before, the contribution of the composition term to the entropy excess can be neglected in the case of WD crystallization (Appendix~\ref{sec:appendix}), so that Equation \eqref{eq_Fc} for the convective heat flux applies in this case as well. Using Equation \eqref{eq_vc2} for the convective velocity, we find 

\begin{equation}\label{eq_del_rad_del_2}
    \nabla_{\mathrm{rad}}-\nabla=a_{0}A(\nabla-\nabla_{e})(\nabla-\nabla_{e}+\nabla_{\mathrm{com}})^{1/2}~,
\end{equation}
which replaces Equation \eqref{eq_del_rad_del}.
Note that from Equation~\eqref{eq_del_rad_del_2}, compositionally driven convection in a subadiabatic thermal background ($\nabla < \nabla_{e}$) has a heat transport directed inward (as in F23).  

Combining Equations~\eqref{eq_gamma}, \eqref{eq_gamma_3}, and \eqref{eq_del_rad_del_2}, we find the fourth-degree equation

\begin{eqnarray}\label{eq_poly_4degree}
    a_{0}\Gamma^{4}+\Gamma^{3}+(1-A^{2}a_{0}\nabla_{\mathrm{com}})\Gamma^{2}&&\nonumber\\    
    -A^{2}(\nabla_{\mathrm{rad}}-\nabla_{\mathrm{ad}}+\nabla_{\mathrm{com}})\Gamma-A^{2}\nabla_{\mathrm{com}}&=&0~,
\end{eqnarray}
which reduces to Equation \eqref{eq_poly} when $\nabla_\mathrm{com}=0$. Unlike Equation \eqref{eq_poly}, Equation \eqref{eq_poly_4degree} does not always have a unique, real, and positive root for given values of $\nabla_{\mathrm{rad}}$, $\nabla_{\mathrm{ad}}$, and $\nabla_{\mathrm{com}}$. However, even though composition gradients are usually incorporated into MLT by using $\nabla_\mathrm{com}$, a more natural formulation is to use the composition flux $F_{X}=\rho v_{c} X \nabla_{X} (\ell/2H_{P})$ as an input parameter rather than composition gradient\footnote{Here, for WD crystallization, we consider a two-species mixture, with abundances $X$ of carbon and $Y=1-X$ of oxygen; the composition gradient is then  $\nabla_{\mathrm{com}}=(\chi_{X}/\chi_{T})\nabla_{X}$.}. The reason for this is that, in Henyey-type codes, the heat and composition fluxes are fundamental variables in the energy and mass conservation equations, so we generally need to calculate the temperature and composition gradients given the fluxes and not the other way around. In the case of white dwarf crystallization, the light element flux is set by the cooling rate of the core.

F23 showed that a natural dimensionless parameter to use that measures the composition flux is 

\begin{equation}\label{eq_tau}
    \tau=F_{X}\frac{H_{p}\chi_{X}}{\rho\kappa_{T}\chi_{T}X\nabla_{\mathrm{ad}}}=a_{0}\mathrm{\Gamma}\frac{\nabla_{\mathrm{com}}}{\nabla_{\mathrm{ad}}}~.
\end{equation}
The composition flux that gives $\tau=1$ has an associated convective heat flux that is equal to the conductive heat flux down the adiabat. Thus, $\tau$ gives a measure of how much the convection tries to adjust the temperature gradient. For their case of zero net heat flux ($\nabla_\mathrm{rad}=0$), F23 found a unique, positive and real solution for $\Gamma$ as a function of $\tau$, with a rapid transition between the regimes of slow convection with $\Gamma\ll 1$ and fast convection with $\Gamma\gg 1$
occurring at $\tau\sim 1$. 

Rewriting Equation \eqref{eq_poly_4degree} in terms of $\tau$ rather than $\nabla_\mathrm{com}$ gives 

\begin{eqnarray}\label{eq_poly_5degree}
    a_{0}\Gamma^{5}+\Gamma^{4}+\Gamma^{3}-A^{2}[\nabla_{\mathrm{rad}}-(1-\tau)\nabla_{\mathrm{ad}}]\Gamma^{2}&&\nonumber \\
    -\frac{A^{2}}{a_{0}}\tau\nabla_{\mathrm{ad}}\Gamma-\frac{A^{2}}{a_{0}}\tau\nabla_{\mathrm{ad}}&=&0~,
\end{eqnarray}
which generalizes the results of F23 to give $\Gamma$ as a function of the two fluxes $\nabla_\mathrm{rad}$ and $\tau$ (setting $\nabla_\mathrm{rad}=0$, Equation~\eqref{eq_poly_5degree} reduces to Equation~(25) of F23).
Similar quartic and quintic equations were derived by \cite{Umezu1988} for overturning convection in massive stars (compare their Equations (25) and (28)). It is straightforward to verify that the same relations between the gradients and $\Gamma$ still hold as we did previously, so that once a solution for $\Gamma$ is obtained, we can find the gradients $\nabla$, $\nabla_e$, and $\nabla_\mathrm{com}$ using Equations \eqref{eq_del}, \eqref{eq_del_del_e}, and \eqref{eq_tau}. We explore the solutions to this equation in the next section.

\subsection{Transition from Efficient to Inefficient Convection}\label{sec_ineff_eff_conv}
In the following, we show that thermohaline convection can be understood as inefficient compositional convection ($\Gamma\ll 1$). Recall that continued convection (either compositional overturning convection or thermohaline convection) demands condition~\eqref{eq_continuous_conv} is satisfied. For the particular case of thermohaline convection, where $\nabla-\nabla_{\mathrm{ad}}+\nabla_{\mathrm{com}}<0$ (Ledoux stable), Equation~\eqref{eq_continuous_conv} requires that both the thermal and composition gradients must simultaneously fulfill

\begin{equation}
    \nabla-\nabla_{e}+\nabla_{\mathrm{com}}>0>\nabla-\nabla_{\mathrm{ad}}+\nabla_{\mathrm{com}}.
\end{equation}
Using Equation~\eqref{eq_gamma}, the condition above can be rewritten in terms of the efficiency $\Gamma$
\begin{equation}\label{ineq_ther}
    \frac{\Gamma}{1+\Gamma}(\nabla-\nabla_{\mathrm{ad}})+\nabla_{\mathrm{com}}>0>(\nabla-\nabla_{\mathrm{ad}})+\nabla_{\mathrm{com}}.
\end{equation}
Since $\nabla_{\mathrm{com}} > 0$ and $\nabla - \nabla_{\mathrm{ad}} < 0$, Equation~\eqref{ineq_ther} requires the convective efficiency to be small $\Gamma \ll 1$ (which is equivalent to having a thermal adjustment time $t_{\mathrm{therm}}$ much smaller than the convective turnover time $t_{\mathrm{conv}}$). This is consistent with the fact that the thermohaline instability develops in a thermally stable background, where thermal diffusion is important.

Now, we show that our formalism gives a mixing coefficient for thermohaline convection that is of the same form as the prescriptions used for thermohaline convection in stellar evolution codes. In the limit $\Gamma \ll 1$, the criterion for continued convection (Equation~\eqref{eq_continuous_conv}) is marginally satisfied. Then, Equation~\eqref{eq_gamma_3} gives $\nabla_{\mathrm{com}}\approx(\nabla_{e}-\nabla)$, and from the middle term of Equation~\eqref{eq_gamma} we can write

\begin{equation}\label{eq:delcom_th}
\nabla_{\mathrm{com}}\approx(\nabla_{\mathrm{ad}}-\nabla)\frac{\Gamma}{1+\Gamma}\approx(\nabla_{\mathrm{ad}}-\nabla)\Gamma.
\end{equation}
If we define the mixing coefficient as $D=v_{c}\ell$ and use the definition of $\Gamma$ in the last term of Equation~\eqref{eq_gamma} we find

\begin{equation}\label{eq:DR_1}
D=2a_{0}\kappa_{T}\left(\frac{\nabla_{\mathrm{com}}}{\nabla_{\mathrm{ad}}-\nabla}\right)=2a_{0}\kappa_{T}R_{0}^{-1},
\end{equation}
where the ratio $R_{0}=(\nabla_{\mathrm{ad}}-\nabla)/\nabla_{\mathrm{com}}$ \citep[as in Equation 43 of][] {Brown2013}. This expression for $D$ is the same as that presented in \citet{Paxton2013} for thermohaline mixing (motivated by the analysis of \citealt{Kippenhahn1980}) if we identify their efficiency parameter $\alpha_{\mathrm{th}}$ as $\alpha_{\mathrm{th}}=4a_{0}/3$, where $a_0$ is our shape parameter. Typical values of the shape parameter $\approx 1/2$--$10$ give values of $\alpha_{\mathrm{th}}$ within the range discussed in \citet{Paxton2013}.
Moreover, if we take $t_{\mathrm{therm}}\approx\ell^{2}/\kappa_{T}4a_{0}$ and $(\ell/H_{p})\nabla_{\mathrm{com}}=-(\varphi/\delta) D\mu/\mu$, we obtain
\begin{equation}\label{v_c_mu}
    v_{\mu}=\frac{-H_{p}}{(\nabla_{\mathrm{ad}}-\nabla)t_{\mathrm{therm}}}\frac{\varphi}{\delta}\frac{D\mu}{\mu},
\end{equation}
the same expression for the thermohaline velocity as \citet{Kato1966}, \citet{Kippenhahn1974}, and \citet{Kippenhahn2013}.

We make clear that our set of equations does not directly provide a solution for the mixing length $\ell$. In Section \ref{sec:WD}, we use the numerical solutions of our MLT equations for $\Gamma$ and the relevant gradients to explore the effect of different assumptions of $\ell$ in the inefficient regime.


\section{The Different Regimes of Compositionally Driven Convection}\label{sec:3}

\begin{figure}
    \centering
    \includegraphics[scale=0.54]{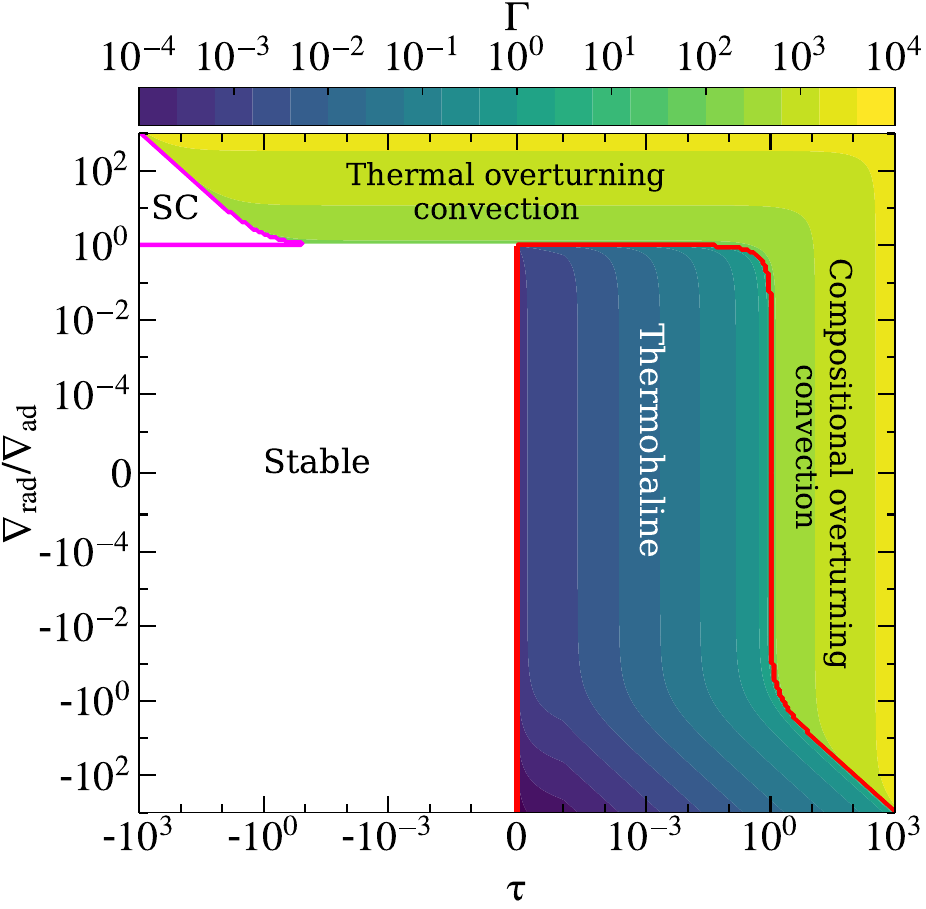}  
    \caption{The convective efficiency $\Gamma$ (Equation~\eqref{eq_gamma}) as a function of the total heat flux as measured by $\nabla_{\mathrm{rad}}/\nabla_{\mathrm{ad}}$ (Equation.~\eqref{eq_Ftot}) and the composition flux as measured by $\tau$ (Equation~\eqref{eq_tau}). The solutions were obtained by solving Equation~\eqref{eq_poly_5degree} numerically. We show the different regimes of convection, as well as the stable region where convection does not occur. The red line shows the transition to the thermohaline convection regime and the magenta line delimits the regime of semiconvection. To solve Equation~\eqref{eq_poly_5degree}, we used $a_{0}=9/4$, $A=2.5\times{10^{4}}$, and $\nabla_{\mathrm{ad}}=1/3$.}
    \label{fig:gamma}
\end{figure}

\begin{figure*}
    \centering
    \includegraphics[width=0.48\textwidth]{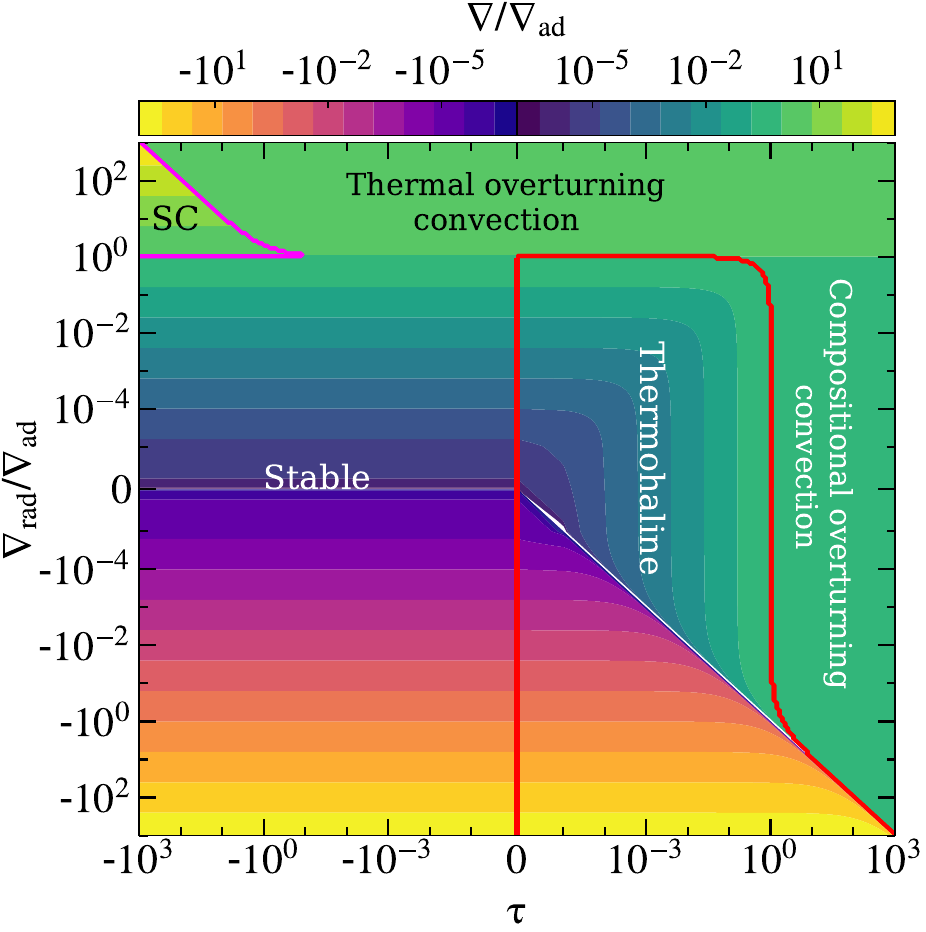}\hspace{0.1in}
    \includegraphics[width=0.48\textwidth]{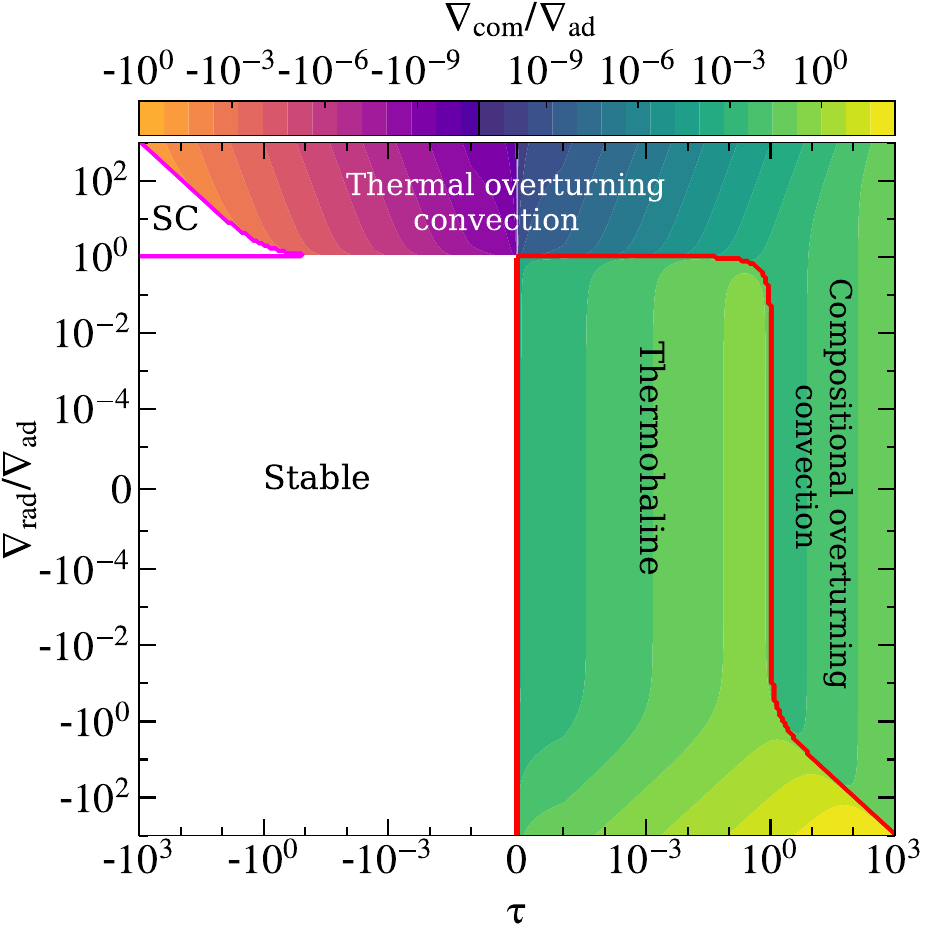}
  \caption{Normalized temperature and composition gradients $\nabla/\nabla_{\mathrm{ad}}$ (left panel) and $\nabla_{\mathrm{com}}/\nabla_{\mathrm{ad}}$ (right panel) as a function of $\nabla_{\mathrm{rad}}/\nabla_{\mathrm{ad}}$ and $\tau$. The solutions were obtained using equations \eqref{eq_del} and \eqref{eq_tau},  and the $\Gamma$ values from Figure \ref{fig:gamma}. In regions that are stable to convection, we set $\Gamma=0$ and $\nabla=\nabla_{\mathrm{rad}}$. 
  }
    \label{fig:nablas}
\end{figure*}

We now use the formalism established in the previous section to investigate the different regimes of compositionally driven convection. We assume that the heat flux and composition flux are specified (as dimensionless parameters $\nabla_\mathrm{rad}/\nabla_\mathrm{ad}$ and $\tau$ respectively; see Equations~\eqref{eq_Ftot} and \eqref{eq_tau}) and calculate the convective efficiency $\Gamma$ (related to the convective velocity and Peclet number as in Equation~\eqref{eq_gamma}) and the temperature and composition gradients $\nabla$ and $\nabla_\mathrm{com}$.

Figure~\ref{fig:gamma} displays the numerical solutions of Equation~\eqref{eq_poly_5degree} for the choice of parameters\footnote{The value of $A$ is chosen to clearly visualize the transition between thermohaline and overturning convection. Realistic values for white dwarfs are much larger ($A\sim 10^{11}$; see Section~\ref{sec:WD}), and the transition is then much sharper (e.g.,~compare Figure~\ref{fig:WD_CO}).} $a_{0}=9/4$, $A=2.5\times{10^{4}}$, and $\nabla_{\mathrm{ad}}=1/3$. Except for situations where the condition $D\rho<0$ (Equation~\eqref{eq_continuous_conv}) was not fulfilled (i.e., regions where convection does not occur and $\Gamma$ is set to 0), there is a unique solution for $\Gamma(\nabla_\mathrm{rad},\tau)$ in almost the entire parameter space. There is a second solution in the top left region of Figure~\ref{fig:gamma} ($\tau < 0$ and $\nabla_{\mathrm{rad}}/\nabla_{\mathrm{ad}} > 0$), but it shows an unphysical discontinuity across $\tau=0$ (see Appendix \ref{ApB}), and so we do not show it here.

The different regimes of convection are labeled in Figure~\ref{fig:gamma}. Whereas we usually think of the boundaries between different convective regimes in terms of the gradients (for example, whether or not the gradients satisfy the Ledoux criterion), here we have parameterized the convection by the heat and composition fluxes. In this case, it is helpful to write Equation \eqref{eq_continuous_conv} for $D\rho<0$ in terms of the fluxes. Replacing $\nabla-\nabla_{e}$ and $\nabla_{\mathrm{com}}$ from Equations \eqref{eq_del_del_e} and \eqref{eq_tau}, and multiplying by $a_{0}\Gamma$ we can write Equation \eqref{eq_continuous_conv} as follows:

\begin{equation}
(\nabla_{\mathrm{rad}}-\nabla_{\mathrm{ad}})(1-\zeta)\frac{a_{0}\Gamma^{2}}{1+\Gamma}+\tau\nabla_{\mathrm{ad}}>0.
\end{equation}
Finally, dividing by $\nabla_{\mathrm{ad}}$ and using the definition of $\zeta$ from Eq.~\eqref{eq:zeta} to write $(1-\zeta)=(1+\Gamma)/[1+\Gamma(1+a_{0}
\Gamma)]=(1+\Gamma)\zeta/a_{0}\Gamma^{2}$, the instability condition in terms of the fluxes is
\begin{equation}\label{eq_drho_fluxes}
    \left(\frac{\nabla_{\mathrm{rad}}}{\nabla_{\mathrm{ad}}}-1\right)\zeta+\tau>0~.
\end{equation}
Since $\zeta$ ranges from $0$ to $1$, we see that a total heat flux $\nabla_\mathrm{rad}>\nabla_\mathrm{ad}$ or a positive composition flux $\tau>0$ destabilize, whereas a total heat flux $\nabla_\mathrm{rad}<\nabla_\mathrm{ad}$ or negative composition flux $\tau<0$ stabilize.

When either $\nabla_\mathrm{rad}$ or $\tau$ are large and positive, we have overturning convection (Ledoux unstable) with $\Gamma\gg 1$. The red curve in Figure~\ref{fig:gamma} bounds the region of thermohaline convection with $\Gamma\ll 1$, which is unstable according to Equation~\eqref{eq_continuous_conv} ($D\rho<0$) but Ledoux stable ($\nabla-\nabla_{\mathrm{ad}}+\nabla_{\mathrm{com}}<0$). To distinguish between each convective regime, we use Equations \eqref{eq_del}, \eqref{eq_del_del_e}, and \eqref{eq_tau} to obtain numerical solutions for $\nabla$, $\nabla_{e}$, and $\nabla_{\mathrm{com}}$. For example, the red curve is defined by the solutions that satisfy $\nabla-\nabla_{\mathrm{ad}}+\nabla_{\mathrm{com}}=0$, where the bounded zone contains all the solutions that fulfill the conditions for thermohaline convection, i.e., $\nabla-\nabla_{e}+\nabla_{\mathrm{com}}>0>\nabla-\nabla_{\mathrm{ad}}+\nabla_{\mathrm{com}}$.

At the top left, we see that overturning convection with $\nabla_\mathrm{rad}>\nabla_\mathrm{ad}$ shuts off if there is a large enough stabilizing composition flux (negative $\tau$). This is the region in which semiconvection would occur (bordered by the magenta curve in Figure~\ref{fig:gamma}, which we draw similarly to the red curve for the thermohaline zone). Note, however, that as semiconvection is not included in our theory, Equation \eqref{eq_continuous_conv} gives $D\rho>0$ in this region and $\Gamma=0$. When both $\tau<0$ and $\nabla_\mathrm{rad}<\nabla_\mathrm{ad}$ (lower left), convection shuts off completely and $\Gamma=0$.

\begin{figure}
    \centering
    \includegraphics[width=0.472\textwidth]{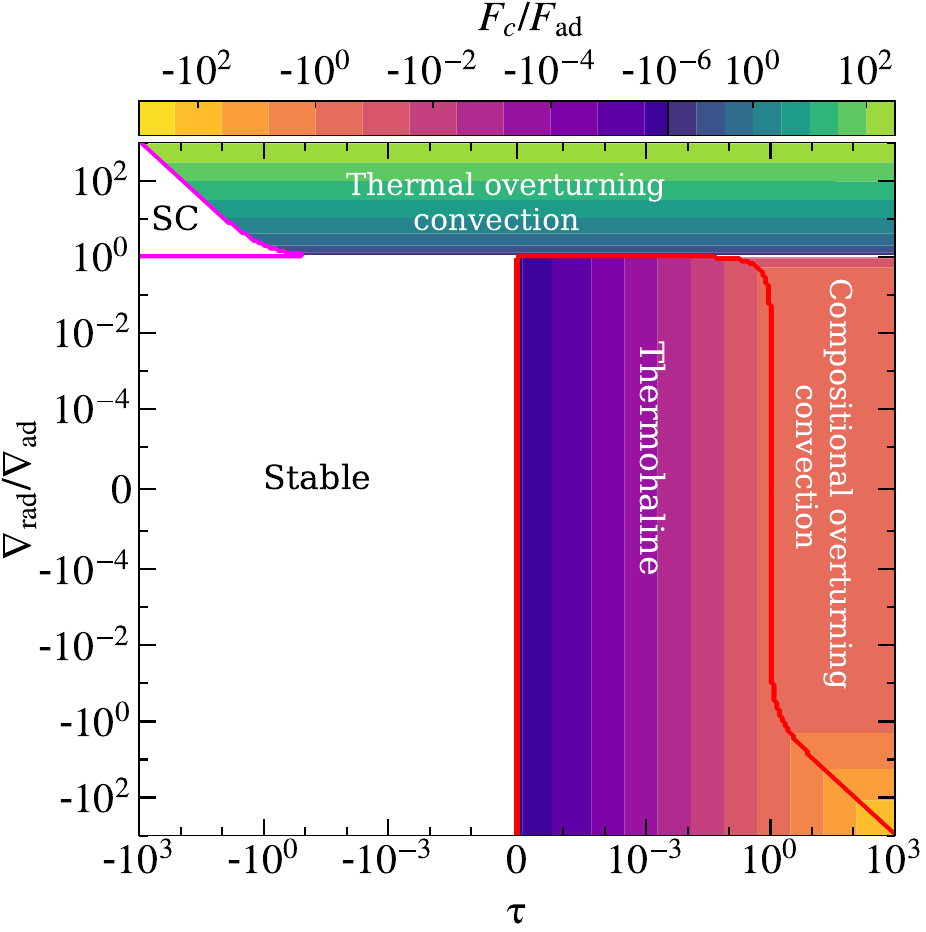}
  \caption{Convective heat flux $F_{c}$ (normalized by $F_{\mathrm{ad}}=\rho c_{P} \kappa_{T} T \nabla_{\mathrm{ad}}/H_{P}$) as a function of $\nabla_{\mathrm{rad}}/\nabla_{\mathrm{ad}}$ and $\tau$. These values were obtained using Equations \eqref{eq_gamma_3} and \eqref{eq_del_rad_del_2}, and the $\Gamma$ values from Figure \ref{fig:gamma}.} 
    \label{fig:nablas2}
\end{figure}

Figure~\ref{fig:gamma} shows that compositionally driven convection has two distinct regimes: slow thermohaline convection with $\Gamma \ll 1$ and fast efficient overturning convection with $\Gamma \gg 1$. As previously found by F23, the transition from slow to fast is extremely rapid and occurs at $\tau\approx 1$ as long as the total heat flux is not too negative ($\nabla_{\mathrm{rad}}>-\nabla_{\mathrm{ad}}$). If $\nabla_{\mathrm{rad}}<-\nabla_{\mathrm{ad}}$, the total inward heat flux, which is stabilizing, can be large enough to compensate for the destabilizing outward composition flux, reducing $\Gamma$. This moves the slow-fast transition to larger values of $\tau\approx -\nabla_\mathrm{rad}/\nabla_\mathrm{ad}$ (lower right of Figure~\ref{fig:gamma}).

Figure~\ref{fig:nablas} shows the behaviour of the gradients $\nabla$ (left panel) and $\nabla_\mathrm{com}$ (right panel) in the different regimes. In the stable regions where $\Gamma=0$ (see Figure~\ref{fig:gamma}), $\nabla$ is just equal to $\nabla_{\mathrm{rad}}$, meaning that the energy transport occurs only by radiation. In regions where $\Gamma\gg 1$, $\nabla/\nabla_{\mathrm{ad}}\rightarrow{1}$, independently of the nature of the dominant destabilizing factor, i.e., outward composition or heat flux. For the thermohaline regime, the value of $\nabla$ depends on both fluxes, increasing from  $\nabla_{\mathrm{rad}}$ to $\sim\nabla_{\mathrm{ad}}$ in the range of $\tau \sim 0$--1. The transition becomes more abrupt as the stabilizing heat flux is more negative. In the thermohaline regime with large inward heat flux, $\nabla$ changes sign. The transition from positive to negative can be understood by setting $\nabla=0$ in Equation \eqref{eq_del} and using the $\Gamma\ll 1$ solution of Equation~\eqref{eq_poly_5degree}, $\tau\approx \zeta(1-\nabla_{\mathrm{rad}}/\nabla_{\mathrm{ad}})$, which gives $\tau=-\nabla_{\mathrm{rad}}/\nabla_{\mathrm{ad}}$ for $\nabla=0$. This agrees with the location of the change in sign of $\nabla$ in the left panel of Figure~\ref{fig:nablas}. 

The right panel of Figure \ref{fig:nablas} shows the solutions for $\nabla_{\mathrm{com}}$. Its sign is given by the direction of the composition flux $\tau$, and its magnitude also is dominated by how large or small is $\tau$. While $\nabla_{\mathrm{com}}$ is generally an increasing function of $\tau$, there is a sudden drop in its magnitude when convection transitions from slow to fast convection. Overall, the composition gradient depends mostly on $\tau$ and is more or less independent of $\nabla_\mathrm{rad}$. The exception is for strong inward heat flux $\nabla_\mathrm{rad}<-\nabla_\mathrm{ad}$, where the composition gradient increases as the inward heat flux becomes stronger at fixed $\tau$. 

Figure \ref{fig:nablas2} shows the convective heat flux $F_{c}$, normalized by $F_{\mathrm{ad}}=\rho c_{P} \kappa_{T} T \nabla_{\mathrm{ad}}/H_{P}$, the conductive heat flux along the adiabat (this is equivalent to normalizing Equation~\eqref{eq_del_rad_del_2} for $\nabla_\mathrm{rad}-\nabla$ by $\nabla_{\mathrm{ad}}$). For thermal overturning convection when $\nabla_\mathrm{rad}>\nabla_\mathrm{ad}$, the convective heat flux is outward, whereas for compositionally driven convection, the convective heat flux is inward. This follows from the fact that $F_c\propto (\nabla-\nabla_{e})$, so it is positive when the regime is superadiabatic and negative for subadiabatic. For overturning convection, $F_c$ depends mainly on $\nabla_\mathrm{rad}$, and for thermohaline convection, $F_c$ depends mainly on $\tau$. 

\section{Application to White Dwarf Crystallization}
\label{sec:WD}

In F23, we used models of cooling white dwarfs from the MESA code to estimate the parameters of convection during the crystallization phase. However, those models did not include phase separation and the convection theory assumed that the conductive and convective heat fluxes exactly balanced, giving $\nabla_\mathrm{rad}=0$. In this section, we improve on those estimates by using cooling models that include phase separation, following \citet{Bauer2023}, and using the total heat flux and composition flux extracted from the model to determine the convective regime and obtain the parameters of convection. We note that, in light of the mixing length theory presented in Section~\ref{sec:new_mlt}, a self-consistent calculation of $\tau$ and $\Gamma$ requires implementing such formalism into 1D evolution models. This will be accomplished in future work.

\begin{figure}
    \centering
    \includegraphics[scale=0.55]{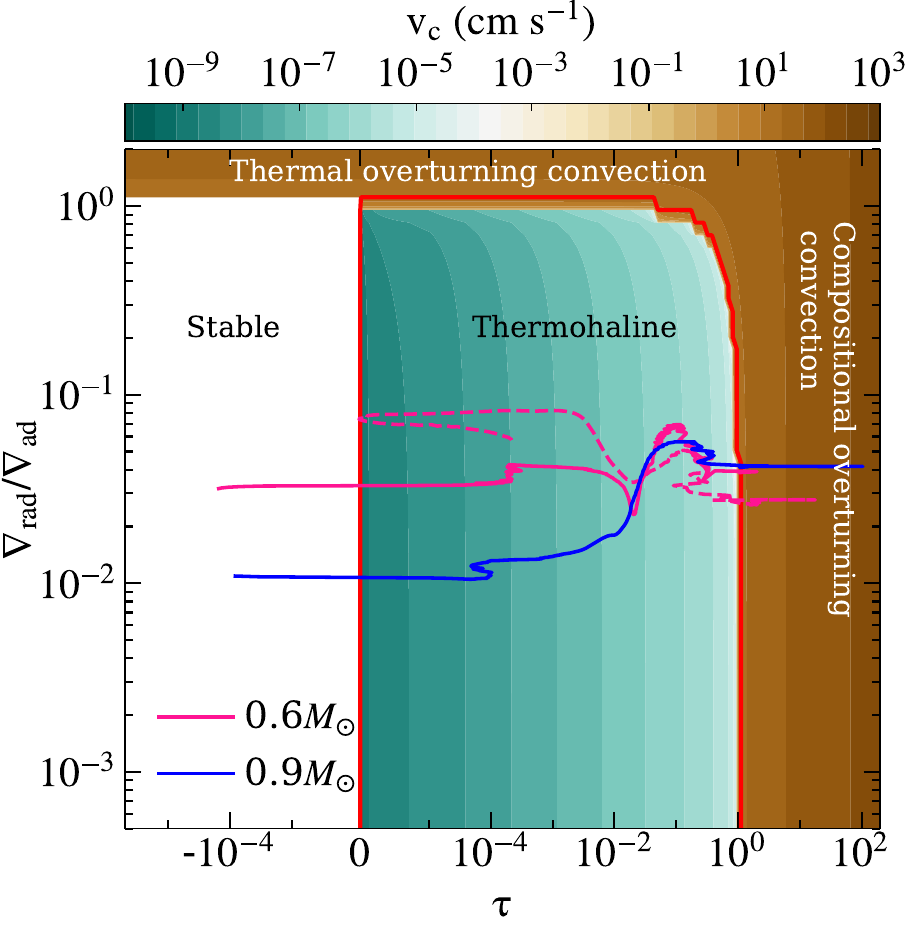} 
  \caption{Evolution tracks in the $\tau$-$\nabla_{\mathrm{rad}}/\nabla_{\mathrm{ad}}$ parameter space for three white dwarf cooling models. The solid lines indicate the models with initial abundances from stellar evolution, while the dashed line denotes the model with a uniform 50/50 C/O core abundance. In this diagram, the star moves from right to left as it evolves. The background color map indicates the magnitude of the convective velocity $v_{c}$ computed from $\Gamma$. We use $a_{0}=9/4$, $A=10^{11}$,  $\nabla_{\mathrm{ad}}=1/3$, $\kappa_{T}=50\ \mathrm{cm^{2}\ s^{-1}}$, and $\ell=H_{P}=10^{8}\ \mathrm{cm}$.}
    \label{fig:WD_CO}
\end{figure}

\begin{figure*}
    \centering
    \includegraphics[width=0.97\textwidth]{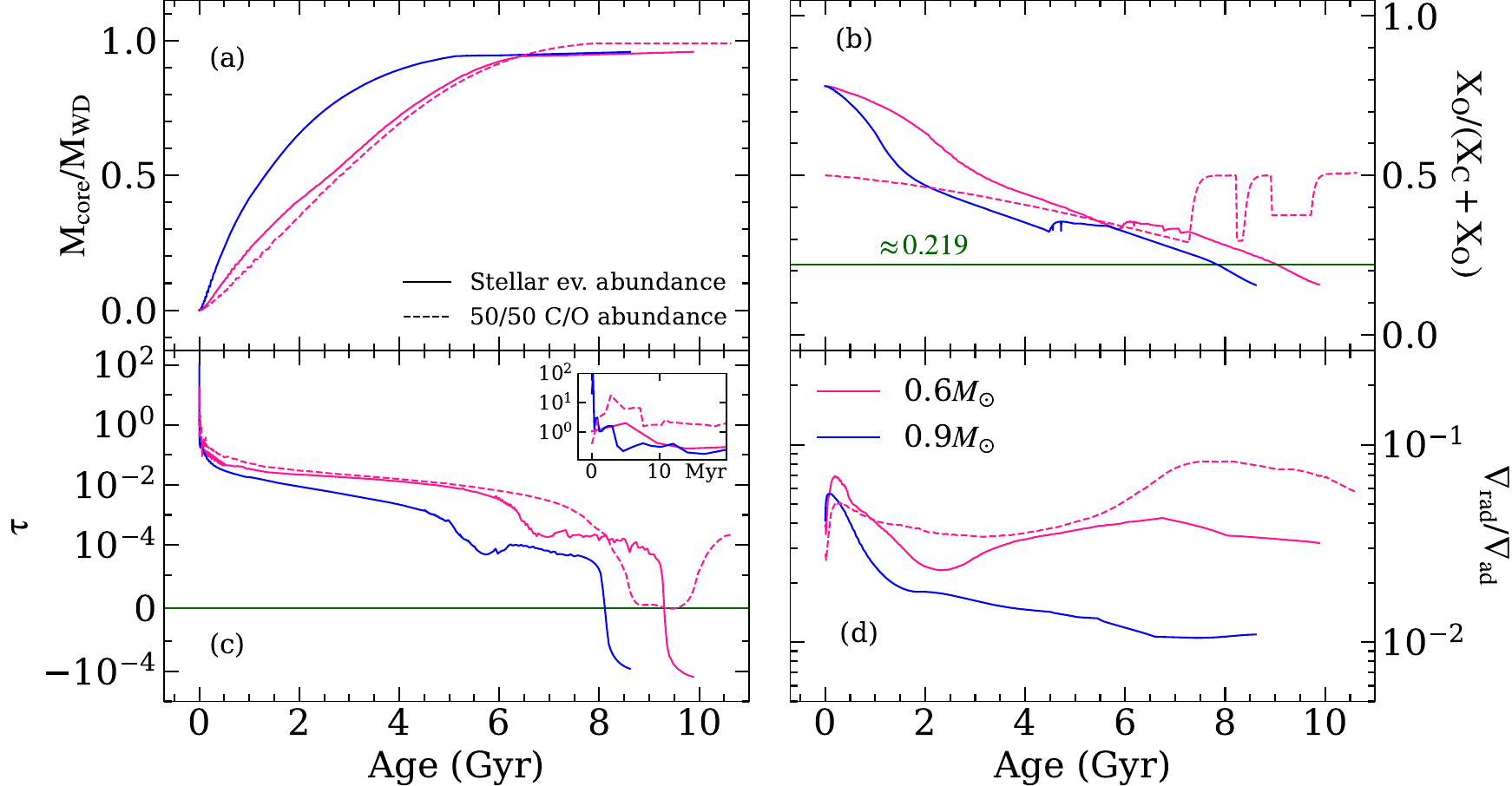} 
  \caption{Time evolution of the convective parameters above the solid-liquid interface following the onset of crystallization for the models shown in Figure~\ref{fig:WD_CO}. To follow the crystallization front and the flux of light elements out of the solid core, we used Equation~\eqref{eq_FX}, the output of the Skye EOS of \citet{Jermyn2021}, the phase diagram of \citet{Blouin2020}, and the phase separation routine in \citet{Bauer2023} (see text for details).}
    \label{fig:WD_params}
\end{figure*}

We used MESA version r23.05.1 \citep{Paxton2011,Paxton2013,Paxton2015,Paxton2018,Paxton2019,Jermyn2023} to run cooling models\footnote{The inlists used in this work are publicly available at \dataset[doi:10.5281/zenodo.10558015]{https://doi.org/10.5281/zenodo.10558015}.} of white dwarfs incorporating phase separation of the C/O core following \citet{Bauer2023}. We adopt two different composition profiles: (1) a realistic profile from the stellar evolution of a 0.6$M_{\odot}$ WD obtained directly from the settled model of the test suite \texttt{wd\_cool\_0.6M}, and (2) an initial profile from \cite{Bauer2023} corresponding to a WD with a uniform core abundance of 50\% carbon (C) and 50\% oxygen (O).  To simplify the calculation, we use a nuclear network containing only C, O, He, and H (with nuclear reactions turned off), so that the internal structure of the star is composed of a C/O core surrounded by a He/H atmosphere. The phase separation calculation takes into account the gravitational energy released due to the redistribution of elements \citep{Bauer2023} and the latent heat released during the phase transition is obtained from the Skye EOS in \citet{Jermyn2021}.

To calculate the flux of light elements we first identified the location of the crystallization front in each model profile. Following the calculation of \citet{Bauer2023}, we use the phase value $\phi=0.9$ from the Skye EOS \citep{Jermyn2021} to set the solid core boundary ($\phi=0$ represents the liquid phase, while $\phi=1$ is the solid phase).

Then, following F23, we compute the composition flux as

\begin{equation}\label{eq_FX}
F_{X}=\dfrac{\dot{M}_{c}\Delta{X}_{\mathrm{melt}}}{4\pi R_{c}^{2}}~,
\end{equation}
where $\dot{M}_{c}$ is the rate of growth of the core mass, $R_{c}$ is the radius of the core, and $\Delta{X}_{\mathrm{melt}}$ is the increase of C in the liquid phase relative to the solid, which is obtained from the liquid-solid composition difference given by the C/O phase diagram of \citet{Blouin2020} and \citet{Blouin2021}. Once we estimate $F_{X}$, we compute the normalized composition flux $\tau$ from the middle term in Equation~\eqref{eq_tau}, using the output profiles of the models to estimate the thermodynamic properties ($\rho,~X,~\chi_T,~\chi_X,~\nabla_{\mathrm{ad}}$). To ensure that these parameters correspond to the liquid just outside the crystallization front in the convection zone, they were taken in the model profiles at the location of a phase value $\phi=0.01$. The local value of $\nabla_{\mathrm{rad}}$ at $\phi=0.01$ representing the total heat flux is also taken from the output profiles. We checked that varying this value of $\phi$ in the range $10^{-4}$ to $0.5$ does not significantly change the results.

The results are shown in Figures \ref{fig:WD_CO} and \ref{fig:WD_params} for two different white dwarf masses, 0.6 and $0.9\ M_\odot$, with the stellar evolution abundance profile. We also show a $0.6\ M_\odot$ model with the 50/50 C/O profile to show the effect of changing the abundance profile. Figure \ref{fig:WD_CO} shows the evolution of the heat and composition fluxes in the $\tau$-- $\nabla_{\mathrm{rad}}/\nabla_{\mathrm{ad}}$ plane. The color map shows the corresponding convective velocity $v_{c}$ obtained from the efficiency parameter $\Gamma$ using Equation~\eqref{eq_gamma}, assuming $a_{0}=9/4$ and taking typical values for $A$, $\nabla_\mathrm{ad}$, and $H_P$ along each trajectory. 
Figure \ref{fig:WD_params} shows the time evolution of the solid core mass, composition at the liquid-solid interface, and the composition and heat fluxes.
We find in all models, that for most of the evolution of the star, convection is in the slow thermohaline regime, where $v_{c}\lesssim 10^{-5}\ \mathrm{cm\ s^{-1}}$, in agreement with F23. However, there is a short phase ($\lesssim 10$ Myr) of fast overturning convection ($\tau>10$) at the beginning of crystallization, with convective velocities $v_{c}\sim 10-100\ \mathrm{cm\ s^{-1}}$ (see the inset in panel (c) of Fig.~\ref{fig:WD_params}). The large composition flux at early times is due to the larger rate of crystallization in the first stages of the cooling (panel (a) of Figure~\ref{fig:WD_params}) and the small core radius at that time.

The value of $\tau$ at the early times following crystallization depends on the temporal resolution used. Smaller core values could be achieved if the initial crystallization time is solved more precisely. In this case, we used a time step of $\sim 10^{5}$ yrs in the early stages for all the models, allowing us to solve an efficient regime that lasts only a few Myr. While we do not attempt to show how small the core could be, in a companion paper, \citet{Fuentes2024}, we checked that using an analytical estimation for $\tau$ also predicts an efficient convection regime that occurs during $\sim10$ Myr from the onset of crystallization.

Panel (d) of Figure \ref{fig:WD_params} shows the total heat flux $\nabla_\mathrm{rad}$ in the liquid just outside the crystallization front. The net heat flux is set by the combination of cooling luminosity, latent heat, and the gravitational energy released during the phase separation. At first, these contributions produce a sudden increase in the total outward heat flux at the onset of crystallization. In the $0.6\ M_\odot$ white dwarf, the latent heat and gravitational energy from phase separation maintain the total heat flux at an approximately constant level as crystallization continues. As expected \citep{Bauer2023}, the more massive $0.9\ M_\odot$ white dwarf cools more quickly, and $\nabla_\mathrm{rad}$ continues to decrease, although at a slower rate than if latent heat and phase separation were not included.

While $\nabla_{\mathrm{rad}}/\nabla_{\mathrm{ad}}$ moves in a tight range of values, keeping approximately constant, the composition flux $\tau$ steadily decreases. At $t\sim 8\ {\rm Gyr}$, we observe an interesting dependence on the abundance profile. For the stellar evolution C/O profile, $\tau$ drops rapidly and becomes negative, whereas $\tau$ becomes small but remains positive for the 50/50 C/O profile. This difference is a direct consequence of the C/O phase diagram. 
 
According to the analytical fit presented in \citet{Blouin2021}, when the number abundance of oxygen in the liquid falls below $x_{\mathrm{O}}^{\ell} \lesssim 0.18$, the contrast between the oxygen abundance in the liquid and solid becomes negative ($\Delta{x_{\mathrm{O}}}<0$). This means that the liquid phase is O-enriched and the solid is enhanced in C, creating a compositionally stable region at the base of the C/O mixture. This transition can be seen in panel (b) of Figure~\ref{fig:WD_params}, where we indicate the normalized O mass fraction $\mathrm{X_{O}/(X_{C}+X_{O}})\approx 0.219$ corresponding to the azeotrope in the phase diagram at $x^{\ell}_{\mathrm{O}}\approx 0.18$. Comparing with panel (c), we see that $\tau$ becomes negative when the O abundance drops below the azeotropic value. In contrast, the normalized mass fraction of oxygen in the models with 50/50 C/O initial abundance never crosses the limit $\mathrm{X_{O}/(X_{C}+X_{O}})\approx 0.219$, and therefore $\tau$ remains positive. This difference is related to the different composition profiles in the two cases. The stellar evolution profile has a central region with uniform abundances surrounded by an outer region in which the oxygen abundance drops (see Figure~8 of \citealt{Bauer2023}). As the convection zone expands into this outer region, the oxygen abundance drops below the azeotropic value. In contrast, in the 50/50 C/O model, the flatter oxygen abundance profile results in a slower decrease in the oxygen abundance.

We have assumed $\ell=H_{p}\approx10^{8}\ \mathrm{cm}$ in all our calculations. While this is likely a reasonable assumption for overturning convection, in the thermohaline regime the mixing length is not well-constrained. We have shown analytically in section \ref{sec_ineff_eff_conv} that, for the limit $\Gamma<<1$ we can reproduce the thermohaline mixing parameter $D=v_{c}\ell$ derived in previous studies \citep[e.g.][]{Kippenhahn1980}.
\citet{MontgomeryDunlap2024} used the thermohaline prescription of \citet{Brown2013} to obtain

\begin{equation}\label{eq:DR_2}
    D\approx C^{2}\nu^{1/2}\kappa_{T}^{1/2}R_{0}^{-1/2},
\end{equation}
where $\nu$ is the kinematic viscosity and $C$ is a constant. They also used a different scaling $D\propto R_{0}^{-0.62}$ which they found gave a better fit to the numerical data.

\begin{figure}
    \centering
    \includegraphics[width=3.3in]{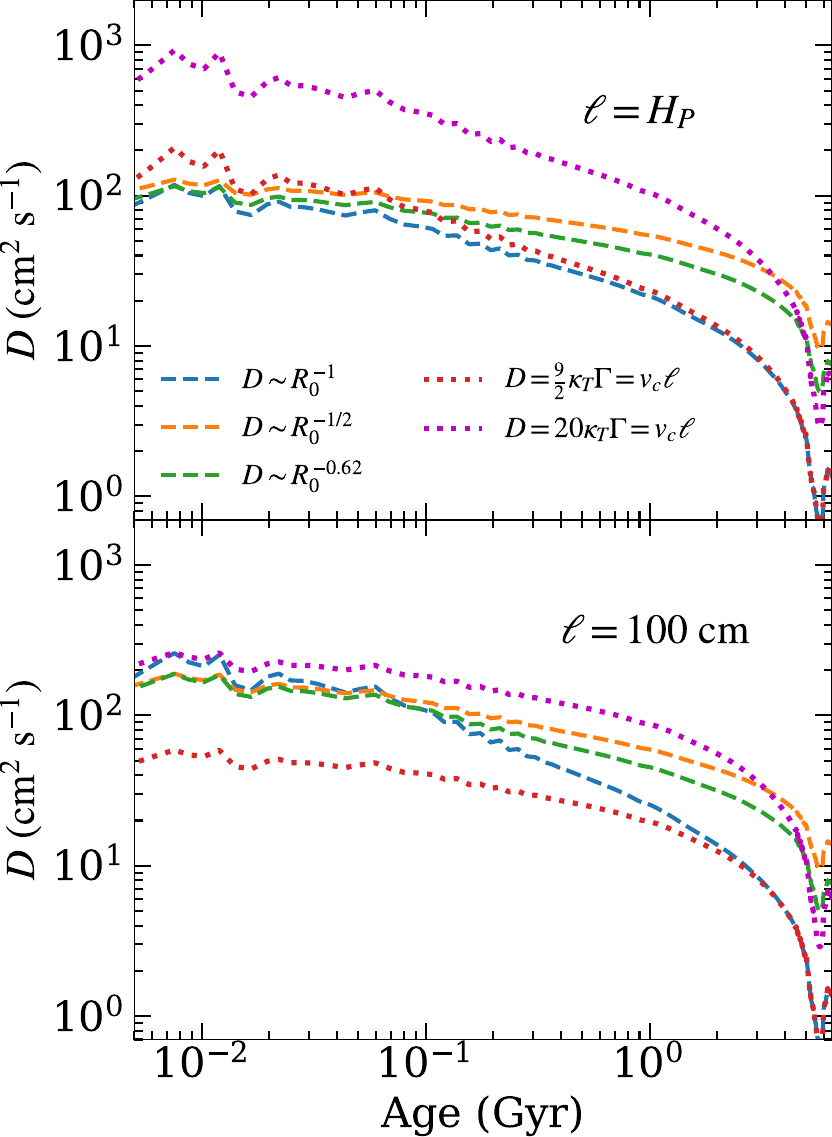}
    \caption{A comparison of different prescriptions for the thermohaline mixing coefficient. We used the stellar parameters and values of $\Gamma$, $\nabla$, and $\nabla_{\mathrm{com}}$ obtained with our MLT for the WD model of 0.9 $M_{\odot}$ to compute the different scaling relations with $R_{0}$. We assumed $a_{0}=9/4$ for the relation shown in Equation~\eqref{eq:DR_1} (blue), and $C=2\pi$ and $\nu=0.4\ \mathrm{cm^{2}s^{-1}}$ for the relations found by \citet{MontgomeryDunlap2024} (orange and green). Additionally, we compare the results with the expected diffusion parameter from $D=v_{c}\ell=2a_{0}\kappa_{T}\Gamma$ for two different values of $a_{0}$ (red and magenta).}
    \label{fig:D}
\end{figure}

These different prescriptions for $D$ are compared in Figure \ref{fig:D} using the solutions found for $\Gamma$, as well as the relevant gradients for the input parameters given by our WD model of 0.9 $M_{\odot}$. To study the dependency on $\ell$ we considered two choices of mixing length: $\ell=H_P$, as in the overturning regime, and $\ell=100\ \mathrm{cm}$ as suggested by \cite{MontgomeryDunlap2024}. Interestingly, we find that the values of $D$ agree within an order of magnitude with the value directly computed from the solution of $\Gamma$ as $D=v_{c}\ell=2a_{0}\kappa_{T}\Gamma$. This may be related to the fact that, in the thermohaline regime, the value of the product $v_c\ell\propto \mathrm{Pe}$ and therefore the mixing coefficient is set directly by the composition flux as shown in F23. Although different prescriptions have different predictions for $v_c$ and $\ell$ individually, their product is determined once the composition flux is specified.

As a consequence of the almost constant thermohaline mixing coefficient, the convective velocity is affected by the mixing length assumption as roughly $v_c\propto \ell^{-1}$, which is why F23 found such a strong enhancement of $v_c$ by rotation, which reduces the effective mixing length. For $\ell=100\ \mathrm{cm}$, the convective velocity in the thermohaline regime is increased by up to six orders of magnitude compared to what is shown in Figure \ref{fig:WD_CO}. Despite this large increase, $v_{c}$ remains $\lesssim 0.5\ \mathrm{cm\ s^{-1}}$ for most of the evolution, since we previously found that $v_{c}$ has typical values $\lesssim 10^{-6}\ \mathrm{cm\ s^{-1}}$ ($\tau\lesssim10^{-1}$ after a few Myr) when assuming $\ell=H_{P}$.

\section{Summary and Conclusions}\label{sec:summary_conclusions}

We have developed a general mixing length theory of convection (MLT) that self-consistently includes the effects of thermal diffusion and composition gradients. Our formulation smoothly transitions between the regimes of fast adiabatic convection at large Peclet number and slow thermohaline convection at low Peclet number. It also allows for both thermally driven and compositionally driven convection, including correctly accounting for the direction of heat transport for compositionally driven convection in a thermally stable background. Once implemented into 1D evolution codes, this will enable investigations of the compositionally driven mixing that occurs in white dwarfs during crystallization. 

Our results improve on existing implementations of MLT in stellar evolution codes. Rather than including the effects of composition by adding composition gradients into the existing MLT equations (with different codes taking different approaches, see discussion in Section~\ref{sec:2.2}), we instead use the heat and composition fluxes as input quantities to derive the resulting convective velocity and thermal and compositional gradients. This leads to a 5th order polynomial (Equation~\eqref{eq_poly_5degree}) that can be solved for the convective efficiency or convective velocity, and from there for the thermal and composition gradients (Equations~\eqref{eq_del}, ~\eqref{eq_del_del_e}, and ~\eqref{eq_tau}). This expression replaces the usual third-order polynomial used in stellar evolution models (Equation~\eqref{eq_poly}). A similar set of equations was found by \citet{Umezu1988, Nakakita1994, Umezu2008, Umezu2009} in the context of core convection in massive stars. 

In stellar evolution codes, our MLT description could be implemented using the heat and composition fluxes obtained from the conservation laws for energy and matter. For example, in the case of white dwarf crystallization, the energy flux depends on the internal energy of the mixture, the latent heat, and the gravitational energy released. The composition flux depends on the correct modeling of the chemical separation during the growth of the crystallized core. Once the composition flux is determined, our MLT prescription can be used to find the corresponding composition gradient for the next time step, similar to the way in which the heat flux is used to determine the temperature gradient.

In this study, we did not attempt to implement our theory in an existing or new code. However, as a first step in applying this theory, we use the output of fiducial WD cooling models from MESA to investigate the regimes of convection that would appear in crystallizing WDs. In doing so, we calculated the heat and composition fluxes above the crystallization front at every stage of the cooling and used them to compute the properties of convection during the star's evolution. We found that, during most of the cooling evolution, convection is in the thermohaline regime, characterized by a small efficiency and convective velocities $v_c\sim 10^{-9}$--$10^{-5}\ \mathrm{cm\ s^{-1}}$. However, at the onset of the crystallization, all the models are found in the efficient regime for a short time ($\lesssim 10\ \mathrm{Myr}$), with much larger values of $v_c \sim 10$--$100\ \mathrm{cm\ s^{-1}}$. This happens because the growth rate of the solid core is large at early times, and its small size also leads to large values of the composition flux (exceeding the dimensionless composition flux $\tau=1$, Equation~\eqref{eq_tau},  which leads to fast convection). 

These results have interesting implications for white dwarf magnetic fields. The efficiency of compositionally driven convection and the corresponding convective velocities play an important role in the crystallization-driven dynamo model proposed to explain highly magnetized WDs \citep{Isern2017,Ginzburg2022}. For a saturated dynamo with $B^{2}/4\pi\sim \rho v_c^{2}$ \citep[e.g.][]{Isern2017}, the magnetic field spans the range $B\sim 10^{-6}$--$10^{5}\rho_{6}^{1/2}\ \mathrm{G}$ for the range of convective velocities that we find above, $v_{c}\sim 10^{-9}$--$10^{2}\ \mathrm{cm\ s^{-1}}$ (assuming a typical density of $10^6\ {\rm g\ cm^{-3}}$). It is important to note that our estimates of the convective velocity do not take into account rotation or magnetic forces. As shown in numerical simulations and scaling laws from mixing length theory, those effects can significantly change the magnitude of the convective velocity \citep[e.g.,][]{Mochkovitch1983, Ginzburg2022, Fuentes2023}. 
However, it is intriguing that the convective velocities we find in the very early phases of crystallization appear to be large enough to generate magnetic fields comparable to those observed. We explore the possibility of a short-lived intense dynamo following the onset of crystallization and the expected magnetic field strengths taking into account rotation in a companion paper \citep{Fuentes2024}

Our calculations can be improved in several respects. The white dwarf cooling models that we use include chemical separation following \cite{Bauer2023}, in which it is assumed that the fluid layer ahead of the crystallization front is fully mixed, and the gravitational energy release from chemical separation is deposited by hand in the model. Implementing the MLT derived here in a time-dependent model would allow to follow the heat and compositional transport in detail, and solve the relevant gradients to construct more consistent temperature and composition profiles during the cooling. 
The onset of crystallization and the short-lived phase of fast convection needs to be followed in more detail, and its dependence on white dwarf mass and the extent of the convection zone during this time determined (which depends on the C/O profile and sets the delay before the field emerges at the white dwarf surface; \citealt{BlatmanGinzburg2024}).
Also, as mentioned above, rotation and magnetic forces should be included when deriving the convective velocity. F23 estimated that rotation will increase the convective velocity significantly in the thermohaline regime (although the convection is still inefficient, with low Peclet number) (see also \citealt{Mochkovitch1983,Ginzburg2022}).

More sophisticated transport models beyond MLT  are needed, in particular in the thermohaline regime. For example, \cite{Brown2013} derived expressions for the fluxes in thermohaline convection based on a series of numerical simulations covering a range of values of the governing parameters. A recent paper by \cite{MontgomeryDunlap2024} used these results (with some modifications) to study mixing in white dwarfs following crystallization. In agreement with F23, they concluded that the convective velocities are much lower than estimated by \cite{Isern1997} and \cite{Ginzburg2022}. However, they also advocated for a much smaller mixing length ($\ell\sim 100\ {\rm cm}$) than we have assumed here ($\ell\approx H_P\sim 10^8\ {\rm cm}$). To investigate this point, we compare the diffusivities from \cite{Brown2013} with the MLT results for both of these choices of $\ell$ in Section \ref{sec:WD}. We find that the mixing coefficients are within an order of magnitude regardless of the chosen prescription or mixing length. The reason for this is that, in the thermohaline regime, the Peclet number $\propto v_c\ell$ is set by the composition flux ($\tau\ll 1$) (as shown by F23), therefore is the effective mixing coefficient which also scales as $v_c\ell$. However, given this result, the convective velocity increases for a smaller $\ell$, scaling as $\ell^{-1}$. This is consistent with the enhancement of $v_c$ found by F23 in the rotating case as the effective mixing length is reduced.

Further multidimensional numerical simulations are also needed to study the transport properties of thermohaline convection under conditions of white dwarf interiors. For example, recent simulations have shown that background magnetic fields can significantly enhance the rate of chemical mixing by fingering convection \citep[e.g.,][]{Harrington_Garaud_2019,Fraser2024}. In this context, it would be also interesting to know how fingers in the thermohaline regime react back on a large intensity field created at the onset of crystallization.

\begin{acknowledgements}
We thank Adrian Fraser for useful conversations and pointing toward important references regarding thermohaline convection. A.C. acknowledges support by NSERC Discovery Grant RGPIN-2023-03620. J.R.F. is supported by NASA through grants 80NSSC19K0267 and 80NSSC20K0193. A.C. M.C.-T. are members of the Centre de Recherche en Astrophysique du Québec (CRAQ) and the Institut Trottier de recherche sur les exoplanètes (iREx).
\end{acknowledgements}

\bibliographystyle{aasjournal}


\appendix
\section{Compositional Corrections to the Convective Heat Flux}\label{sec:appendix}

In Section \ref{sec:2.1}, when we write the convective heat flux as $F_c=\rho v_c TDs$ with $Ds$ being the entropy excess of a convecting fluid element, we take $TDs = c_P DT$, ignoring any contribution to the entropy from composition terms. Here, we calculate the composition terms and show that they make a small correction for white dwarf models. We also discuss the relation between the entropy and enthalpy excess when composition terms are included. 

For simplicity here, we consider a two-component mixture, so we can characterize the composition with a single mass fraction $X$, the mass fraction of the lighter species. Since the fluid element remains in pressure balance with the surroundings $\Delta P=0$, we also take $P$ to be constant, and so

\begin{equation}
    TDs = T\left.{\partial s\over \partial T}\right|_{X,P} DT + T\left.{\partial s\over \partial X}\right|_{T,P} DX.
\end{equation}
Following \cite{MedinCumming2015}, we define $b_X \equiv -X\left.\partial s/\partial X\right|_{T,P}$, and since $T\left.\partial s/\partial T\right|_{X,P} = c_P$, the specific heat capacity at constant pressure, we obtain

\begin{equation}
    Ds=c_{P}\frac{\ell}{2H_{P}}\left(\nabla-\nabla_{e}-\frac{b_{X}}{c_{P}}\nabla_{X}\right),
\end{equation}
and therefore

\begin{equation}\label{eq:Fc_comp}
    F_{c}=\rho v_{c}c_{P}T\frac{\ell}{2H_{P}}\left(\nabla-\nabla_{e}-\frac{b_{X}}{c_{P}}\nabla_{X}\right).
\end{equation}
A similar expression was obtained by \citet{MedinCumming2015} in the adiabatic limit.

\begin{figure}
    \centering
    \includegraphics[width=3.2in]{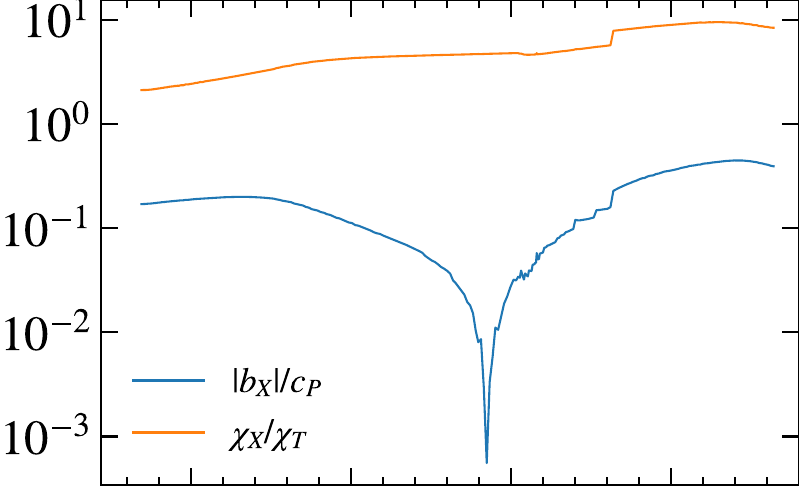}
    \includegraphics[width=3.2in]{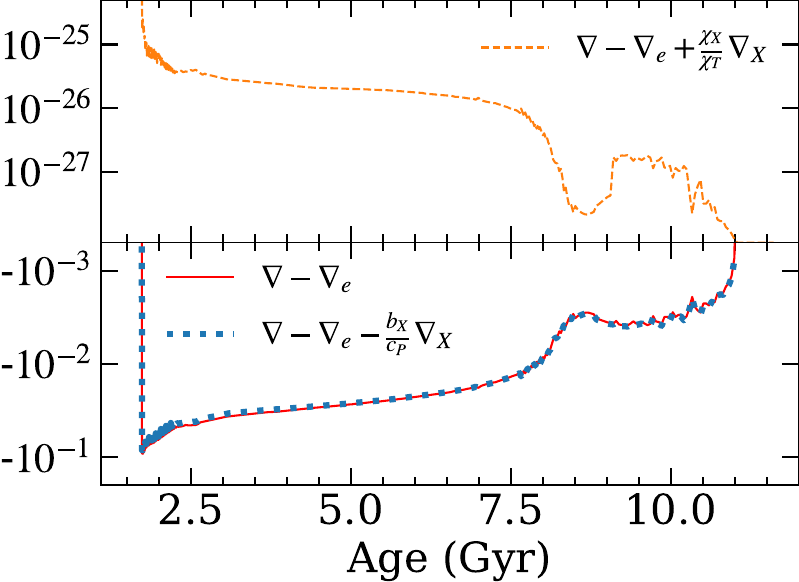} 
  \caption{{\em Top panel:} the coefficients of the composition terms in the entropy ($b_X/c_P$, Equation~\eqref{eq:Fc_comp}) and density ($\chi_X/\chi_T$; Equation~\eqref{eq_del_rho_comp}) excesses $Ds$ and $D\rho$. Values are taken from the 0.6 $M_\odot$ white dwarf model with stellar evolution abundance profile presented in Section \ref{sec:WD}. The ratio $(b_X/c_P)/(\chi_X/\chi_T)\ll 1$ gives the correction to the heat flux from composition terms. {\em Bottom panel:} The solid and dotted curves show the gradient terms in the heat flux without and with composition terms included. The dashed curve shows the gradient terms in $D\rho$ for comparison.}
    \label{fig:chis}
\end{figure}

Figure \ref{fig:chis} shows the value of $b_X/c_P$ and the effect of the composition term on the heat flux for the MESA model of a 0.6 $M_\odot$ white dwarf discussed in Section \ref{sec:WD}. We find that $|b_X|/c_{P}$ takes a value in the range $\approx 1/3$--$2/3$ (the cusp in the profile of $|b_X|$ shown in the top panel of Figure \ref{fig:chis} is because $b_X$ changes sign as the ratio of C/O evolves). However, even though $b_X/c_{P}$ is of order unity, the effect on the heat flux is small, as can be seen by comparing the solid and dotted curves in the lower panel of Figure \ref{fig:chis}. The reason is that, in the convection zone, the instability criterion is only exceeded by a tiny amount, so $\nabla-\nabla_e\approx- (\chi_X/\chi_T)\nabla_X$ (see Equation~\eqref{eq_continuous_conv}). Including the composition term in Equation~\eqref{eq:Fc_comp} therefore leads to a correction $(b_X/c_P)(\chi_T/\chi_X)$ which is $\ll 1$ (since $\chi_X/\chi_T$ is large; top panel of Figure~\ref{fig:chis}). 

As discussed in Section \ref{sec:2.2}, many treatments of MLT with composition gradients are based on writing the heat flux as $F_c=\rho v_c Dh$, where $h$ is the specific enthalpy. Without any composition terms, it is straightforward to show that $Dh=TDs=c_PDT$, so that the enthalpy does in fact measure the heat content and is appropriate to use in the expression for the convective flux. However, the compositional corrections to the enthalpy are very different from the compositional corrections to the entropy. To see this, we can use $c_P=\left.\partial h/\partial T\right|_{X,P}$ to write 

\begin{equation}\label{eq:Dh_appendix}
    Dh = c_P\left(DT - \left.{\partial T\over \partial X}\right|_{h,P}DX\right).
\end{equation}
The density contrast $D\rho$ can be written in a similar way: 

\begin{equation}\label{eq:Drho_appendix}
    D\rho = -{\rho\over T}{\chi_T\over \chi_\rho}\left(DT - \left.{\partial T\over \partial X}\right|_{\rho,P}DX\right).
\end{equation}
For an ideal gas, the enthalpy is proportional to $P/\rho$, and so the $\partial T/\partial X$ derivatives appearing in Equations \eqref{eq:Dh_appendix}) and \eqref{eq:Drho_appendix}) are the same. This means that $Dh\propto D\rho\propto \nabla-\nabla_e+(\chi_X/\chi_T)\nabla_X$, so that assuming $F_c\propto Dh$ then leads to a convective heat flux $F_c\propto v_c(\nabla-\nabla_e+(\chi_X/\chi_T)\nabla_X)$. However, as we see above, the compositional corrections to entropy are actually much smaller than the compositional corrections to the density (which set the instability criterion), so writing the convective heat flux in terms of enthalpy overestimates the compositional terms and gives an incorrect direction of this quantity for the compositionally driven convection (see middle and lower panels of Figure \ref{fig:chis}).

\section{The Other Solution for $\tau<0$ and $\nabla_{\mathrm{rad}}/\nabla_{\mathrm{ad}}>0$}\label{ApB}

\begin{figure}
    \centering
    \includegraphics[scale=0.54]{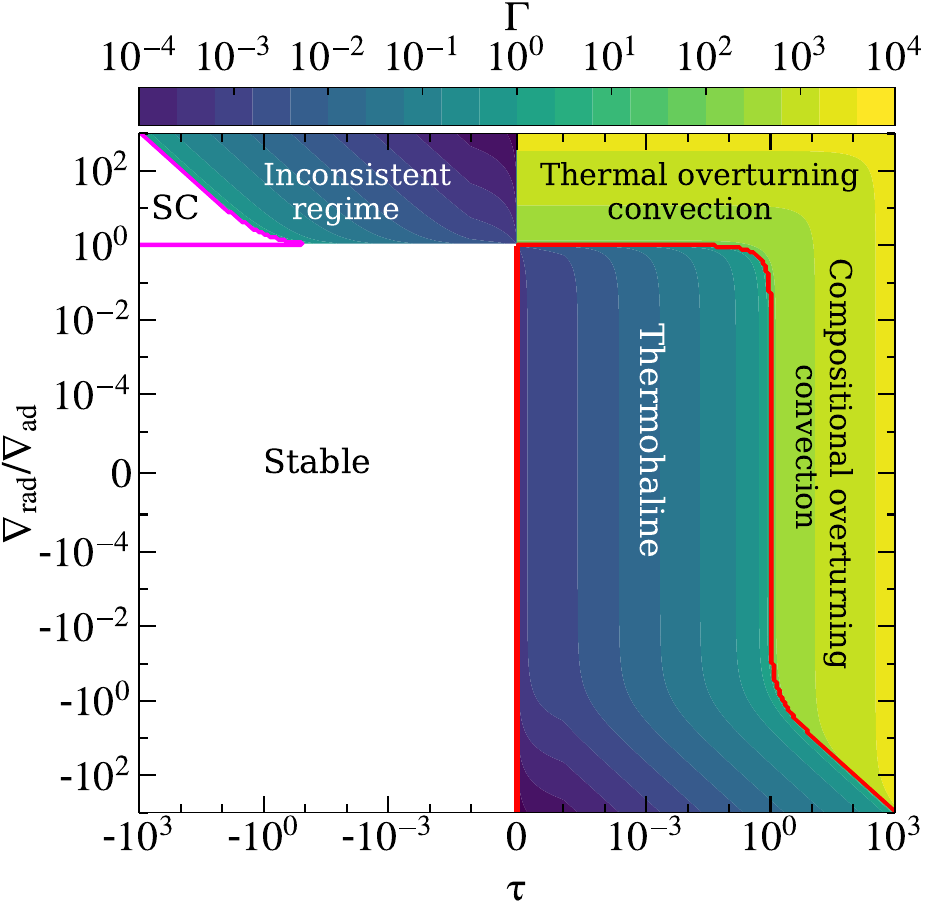}  
    \caption{As Figure~\ref{fig:gamma}, except now showing the second solution to Equation \eqref{eq_poly_5degree} in the region $\tau<0$ and $\nabla_\mathrm{rad}/\nabla_\mathrm{ad}>1$.}
    \label{fig:gamma_no}
\end{figure}

As discussed in Section \ref{sec:3}, the quintic Equation \eqref{eq_poly_5degree} has a second solution in the region of parameter space $\tau<0$ and $\nabla_{\mathrm{rad}}/\nabla_{\mathrm{ad}}>0$. For completeness, we show the second solution in Figure~\ref{fig:gamma_no}, which should be compared with Figure~\ref{fig:gamma}. \citet{Nakakita1994} showed that the second solution is unstable for some choices of mixing length, but were not able to completely rule out that it could be relevant for overturning convection in massive stars (see also \citealt{Umezu2009}). Here, looking broadly across the parameter space, we see that $\Gamma$ changes discontinuously across $\tau=0$ in the overturning convection regime. Since we do not expect the convective velocity to change drastically on going from a small inward composition flux to outward composition flux, we reject this solution as unphysical. In addition, this solution has the property that $\Gamma$ decreases with increasing composition or heat flux, which is opposite to the expected behavior. 

\end{document}